\documentclass[graybox]{svmult}

\usepackage{type1cm}        
\usepackage{makeidx}         
\usepackage{graphicx}        
\usepackage{multicol}        
\usepackage[bottom]{footmisc}

\usepackage{newtxtext}       %
\usepackage{newtxmath}       
\usepackage{array,multirow}
\usepackage[caption=false]{subfig}

\usepackage[final,inline,nomargin]{fixme} \fxsetup{theme=color} 

\FXRegisterAuthor{mz}{amz}{\color{olive}MZ}
\FXRegisterAuthor{cp}{cfp}{\color{blue}CP}
\FXRegisterAuthor{cg}{cgg}{\color{teal}CG}

\makeindex             

\begin{document}

\title*{
    Identifying tax evasion in Mexico with tools from network science and machine learning
}
\titlerunning{Identifying tax evasion with network science and machine learning}

\author{
    Martin Zumaya,
    Rita Guerrero,
    Eduardo Islas,
    Omar Pineda,
    Carlos Gershenson,
    Gerardo Iñiguez, and
    Carlos Pineda
}
\authorrunning{Zumaya, M., \emph{et al.}}

\newcommand{\unam}{Universidad Nacional Aut\'{o}noma de M\'{e}xico, 
01000 Ciudad de M\'{e}xico, Mexico}
\institute{
    Martin Zumaya \at
    Programa Universitario de Estudios sobre Democracia, Justicia y Sociedad
    \&
    Centro de Ciencias de la Complejidad, \unam. \\
    \email{martin.zumaya@puedjs.unam.mx}
    \and
    Rita Guerrero \at
    Plantel Del Valle, Universidad Autónoma de la Ciudad de M\'{e}xico,
    03100 Ciudad de M\'{e}xico, Mexico.
    \and
    Eduardo Islas \at
    Subsecretaría de Fiscalización y Combate a la Corrupción, Secretaría de la Función Pública,
    01020 Ciudad de M\'{e}xico, Mexico. \\
    \and
    Omar Pineda \at
Azure Core Security Services, Microsoft, Redmond, WA 98052, United States; \\
    Posgrado en Ciencia e Ingenier\'ia de la Computaci\'on \& Centro de Ciencias de la Complejidad, \unam \\
    \and
    Carlos Gershenson \at
    Instituto de Investigaciones en Matemáticas Aplicadas y Sistemas \&
    Centro de Ciencias de la Complejidad, \unam; \\
    Lakeside Labs GmbH, 9020 Klagenfurt am Wörthersee, Austria. \\
    \email{cgg@unam.mx}
    \and
    Gerardo Iñiguez \at
    Department of Network and Data Science, Central European University, 1100 Vienna, Austria; \\
    Department of Computer Science, Aalto University School of Science, 00076 Aalto, Finland; \\
    Centro de Ciencias de la Complejidad, \unam. \\
    \email{iniguezg@ceu.edu}
    \and
    Carlos Pineda \at
    Instituto de Física, \unam.\\
}
%
%
\maketitle
\abstract{ 
Mexico has kept electronic records of all taxable transactions since
2014. Anonymized data collected by the Mexican federal government comprises
more than 80 million contributors (individuals and companies) and almost 7 billion
monthly-aggregations of invoices among contributors between January 2015 and
December 2018. This data includes a list of almost ten thousand contributors already
identified as tax evaders, due to their activities fabricating invoices for
non-existing products or services so that recipients can evade taxes.
Harnessing this extensive dataset, we build monthly and yearly temporal
networks where nodes are contributors and directed links are invoices produced
in a given time slice. Exploring the properties of the network neighborhoods
around tax evaders, we show that their interaction patterns differ from those
of the majority of contributors. In particular, invoicing loops between tax
evaders and their clients are over-represented. With this insight, we use two
machine-learning methods to classify other contributors as suspects of tax
evasion: deep neural networks and random forests. We train each method with a
portion of the tax evader list and test it with the rest, obtaining more
than 0.9 accuracy with both methods. By using the complete dataset of
contributors, each method classifies more than 100 thousand suspects of tax evasion,
with more than 40 thousand suspects classified by both methods. We further reduce the
number of suspects by focusing on those with a short network distance from
known tax evaders. We thus obtain a list of highly suspicious contributors
sorted by the amount of evaded tax, valuable information for the authorities to
further investigate illegal tax activity in Mexico. With our methods, we
estimate previously undetected tax evasion in the order of \$10 billion USD per year
by about 10 thousand contributors.  } 

\section{Introduction} \label{sec:introduction}

Tax has a
crucial role in the economic growth and welfare of the general population. Paid
taxes allow for government spending and public expenditures (in the short,
medium, and long terms) such as education, healthcare, housing, pensions,
security, and infrastructure~\cite{CPEUM}. Tax collection in Mexico is
determined by a series of laws ({\it Ley de Ingresos de la Federación}, in
Spanish) specifying the eligibility of taxpayers, tax rates and types, as well
as the periodicity of fees and means of payment.

Despite the general view that taxes are justified since they allow the Mexican state to fund
beneficial activities for society, some individuals, corporations, or trusts may decide to illegally
evade taxes by misrepresenting their state of affairs to the Tax Administration Service in Mexico
({\it Servicio de Administración Tributaria}, or SAT). In this sense, tax evasion is the reduction
of constitutional tax liability by dishonest reporting, such as understating financial gains or
overstating deductions \cite{tovar2000evasion}.

Since 2014, Mexico has kept electronic records of all taxable transactions by
means of a digital receipt or invoice known as {\it Comprobante Fiscal Digital
por Internet} (CFDI). Each of these mandatory receipts includes data on the
product or service transferred between taxpayers, date of transaction, cost,
and corresponding tax amount. CFDIs are XML documents with technical
specifications updated annually by SAT, including a certification seal that can
only be produced by authorized parties \cite{CFDISAP}. Since CFDIs potentially
uncover networks of individuals and legal entities involved in commercial
transactions leading to tax revenue, they are an integral part of formal
investigations by SAT in tax evasion, money laundering, and other tax-related
illegal activities.

Among all forms of tax evasion, here we focus in situations where taxpayers
issue CFDIs in the absence of actual economic activity to increase tax
deductions. Such legal entities, known as `enterprises billing simulated
operations' (`{\it empresas que facturan operaciones simuladas}', or EFOS), are
typically characterized by a lack of employees or infrastructure, as well as a
fake or constantly changing address used to avoid detection. As a result of
investigations already led by SAT, EFOS can be classified as {\it definitive}
or {\it alleged} tax evaders \cite{DefEFO}. In order to operate, EFOS require
the participation of `enterprises deducting simulated operations' (`{\it
empresas que deducen operaciones simuladas}', or EDOS). Even if EDOS engage in
illegal activity alongside EFOS, they tend to have demonstrable stability in
their workforce, assets, and tax contributions. By receiving CFDIs associated
with simulated operations, EDOS aim to reduce their tax rate to avoid payments
to SAT and eventually obtain further fiscal benefits.

In order to decrease the risk of systematic tax evasion, the Mexican federal
government has established fiscal law (known as `{\it Código Fiscal de la
Federación}') describing the official procedure to identify taxpayers as EFOS:
a) First, SAT determines the lack of actual economic activity behind a set of
CFDIs. b) Then, the government notifies the relevant legal entities via its
official publication (`{\it Diario Oficial de la Federación}'). c) Alleged EFOS
have 15 days to contest the claim. d) Associated EDOS (that have received the
suspicious CFDIs) can correct their standing with SAT by resubmitting
appropriate tax forms. e) Finally, if any tax revenue has been lost to illegal
activity, SAT classifies relevant taxpayers as definitive EFOS and specifies
the type of crime following fiscal law. Definitive EFOS are unable to emit
further CFDIs.

Despite its effectiveness, the official procedure of detecting tax evasion is
time and resource consuming, particularly in initial stages of the process
where suspicious CFDIs and alleged EFOS need to be manually selected from
millions of contributors and billions of transactions each year. In order to
complement these efforts, automated computational and statistical techniques
(with tools from network science and machine learning) can be used to
characterize the network of issued/received CFDIs between definitive EFOS and
other taxpayers.  By analyzing the temporal properties of the network of all
taxable transactions in Mexico from 2015 to 2018, here we show evidence of a
group of highly suspicious contributors with behavior statistically similar to
that of definitive EFOS, as well as estimates of previously undetected tax
evasion. When combined with current practices at SAT, this information has the
potential to increase the efficacy of the governmental response to illegal tax
activity in Mexico.

\section{Data} \label{sec:data}

Taxpayers in Mexico are identified by their tax number (`{\it Registro Federal
de Contribuyentes}', or RFC), a unique string of alphanumerical characters used
to emit and receive invoices, submit tax statements and engage in other
procedures. Since 2014, a large number of income and outcome transactions
between taxpayers in Mexico have been recorded in CFDIs and stored by SAT.


The data used in this work includes:
\begin{itemize}
    \item A set of 81,511,015 taxpayer identifiers, anonymized to protect
          individual identities, which we denote RFCAs, and categorical information for
          each one of them that includes: taxpayer type
         (individual or legal entity), location, date of registration, and economical sector and
         activity.
    \item A total of 6,823,415,757 monthly CFDI aggregated emissions between taxpayers distributed between
        January 2015 and December 2018, which include the RFCA of both the emitter and receiver of
        the transaction, the month and year, type (either income or outcome), the number of transactions for that month, and the total amounts
        associated with them.
    \item A list of 8,570 RFCAs previously identified by Mexican government authorities as definitive or alleged
        EFOS. We use this data to train
        machine learning models and as focal points in the network science approach.
\end{itemize}

In the 48 months of CFDI emissions we analyze, 7,571,093 RFCAs emitted at least one CFDI, so that the
already identified evaders (EFOS) account only for 0.0072\% of active taxpayers (those who at
least emitted one receipt during the period of study). This means that the data is highly unbalanced:
the ratio between the identified class (EFOS) and the undetermined ones is quite different, which
has an impact in the design of the models and approaches we use.

In what follows, when we refer to a RFCA as an EFOS, we mean those taxpayers already identified by
the authorities as either definitive or alleged evaders. Unclassified RFCAs correspond to those that
have not been classified as EFOS by the authorities, and we refer as suspects to those RFCA which are
classified as possible evaders by our methods.

\section{Results} \label{sec:results} 

\subsection{Deep Neural Networks} \label{subsec:deppNN}
As a first classification method of unknown RFCs on whether they behave
similarly (or not) to definitive EFOS, we implement an {\it artificial neural network} (ANN).
%
%
ANNs are
models of automatic learning inspired by the human brain. They consist of a
collection of interconnected mathematical functions with characteristics
analogous to those of biological neurons and are thus called neurons.
Just like biological neurons, an artificial neuron
collects and classifies information based on its input connections with other
neurons, and thereby alternates between an active and inactive state.  The
connections between neurons have a weight associated with them representative
of the intensity of the interaction, such that highly weighted connections are
more relevant than those associated with a low weight to modify their
activation state. It is via modification of the connections weight between
neurons that a neural network learns to identify patterns, a process referred
to as {\it training}.

Neurons of an artificial neural network are often divided into different
layers: an input layer that receives the data for classification; hidden layers
that undertake the classification process through modification of the weights
among neurons and adjustment of the input of data weights until the
classification undertaken by the network is optimal; and an output layer,
from which the final result of the networks' classification of input data is
derived. The output of the network is compared with the desired outcome via a
loss function that yields an error quantifier. During training, these errors
are propagated through the network to update the weights and minimize the loss
function. ANNs have been used in a variety of tasks, including computer 
vision~\cite{hongtao2016applications}, voice
recognition~\cite{venayagamoorthy1998voice}, automatic translation
\cite{zhang2015deep},
board and video
games \cite{tesauro1988neural, clark2015training,
NeuralState} and medical diagnostics \cite{amato2013artificial}. They have also
been used in a
variety of applications in financial services, from forecasting to market
studies \cite{kimoto1990stock, mizuno1998application,
wilson1994bankruptcy}, to fraud detection \cite{CreditCardFraud} and risk assessment \cite{trippi1992neural, yu2008credit}. A
neural network can evaluate price data and discover opportunities to make
commercial decisions based on data analysis. Networks can distinguish subtle,
non-linear inter-dependencies and patterns that other methods of technical
analysis cannot.

\subsubsection{Data preparation} 


%

In our implementation, we design an ANN 
that receives input data from all
CFDIs associated with an issuing RFCA. By means of a technique referred to as
{\it re-sampling}~\cite{japkowicz2000class}, we form a balanced sample of unknown RFCAs
and definitive EFOS. The re-sampling method considered in this implementation is comprised
of random re-sampling of the small class (CFDIs issued by definitive EFOS)
until it contains as many examples as the other class, in order to finally
obtain a large dataset with the same quantity of CFDIs issued by unknown RFCAs
and definitive EFOS.

The model associates each RFCA with a value between 0 and 1 related to the
probability that it will be an EFOS. In what follows, we describe the
procedure used to design, train and evaluate the ANN. We will follow by presenting some
results and conclusions.

\subsubsection{Modeling} 
\paragraph{ANN design} 

A {\it dynamic recurrent neuronal network} (DRNN) is a special type of neural
network that allows introduction of an arbitrary number of rows of data (input
variables) at the same time, which is useful in this context since RFCAs have
varying amounts of issued CFDIs. Recurrent neural networks are structures in
which the output of each execution step provide the input to the
following step; this enables them to retain learned information over time. {\it
Long short term memory} (LSTM)\footnote{LSTM cells are a network topology
developed for the first time by Hochreiter and
Schmidhuber~\cite{Hochreiter:1997:LSM} to eliminate the problem of vanishing
gradient~\cite{Hochreiter:1991} through the introduction of a memory mechanism.
A gradient measures how much the output of a function changes if the input
changes a little. The problem is that, for very deep networks, the gradient of
errors dissipates rapidly over time, ending up being very small and this
prevents a change in the weighted values. Networks with this problem are
capable of learning short-term dependencies, but often have difficulty learning
long-term dependencies.} describes the design of artificial neurons, i.e.,
those that give memory to the ANN. These neurons have the best known
performance to date and are particularly effective for datasets derived from
time series~\cite{greff2016lstm, yin2017comparative, chung2014empirical}. In
particular, out of several structures tested, the best performance was obtained
with an DRNN with three LSTM cell layers, each with 256 neurons, using a
hyperbolic tangential function to calculate their internal state\footnote{These
three layers correspond to the hidden layers that classify the input data. In
addition to the hidden layers, the network has an input and an output layer.}. 

\begin{figure} 
	\centering
	\includegraphics[width=0.75\textwidth]{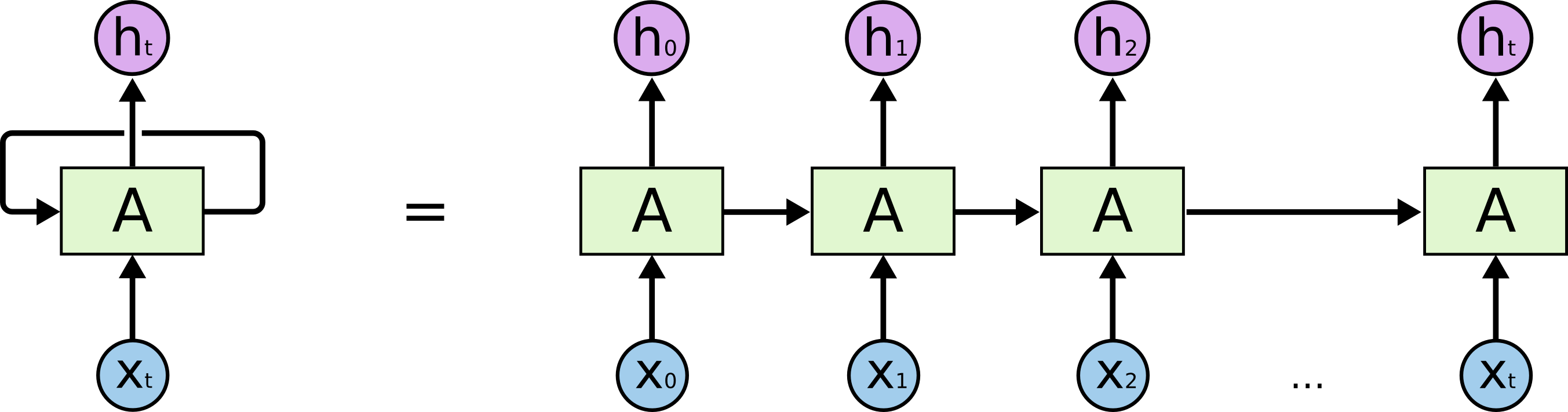}
    \caption{
One part of neural network $A$ observes an input $x_t$ and calculates a value
$h_t$. The cycle allows information to flow from one step of the network to the
next. If we unroll the cycle, one recurrent neural network can be considered as
multiple copies of the same network, such that each one passes a message to a
successor. 
%
    }
\label{fig:RNN}
\end{figure}

It is worth noting that the connections of an DRNN are not only among different
layers, but are also connections from a neuron to itself over time. This means
that the error propagation for weight adjustment occurs, not only among
different nodes, but also between the same node and different time steps, as
shown in Fig.~\ref{fig:RNN}.

An LSTM cell is controlled by three gates: the Forget gate, the Input gate and
the Output gate. Each gate within the cell is a different neural network that
decides what information is allowed in the cell's state, and which functions as
the network's memory. The gates can learn what information is relevant to save
or forget during the exercise. The Forget gate controls the amount of
information that will be stored in the memory and discards irrelevant
information. The Input gate controls the amount of new input that will be
stored in the memory, i.e., determines the importance of new information.
Finally, the Output gateway determines the characteristics of the analyzed
information to obtain an output that will allow correct classification. \par

The architecture of the neuronal network used to classify RFCA as possible
EFOS
is made up of three hidden LSTM cell layers each with 256 neurons connecting
each neuron in a layer to a neuron in the following layer. The network unrolls
over time to analyze all the invoices issued by an RFCA and, from what has been
analyzed, classifies it into EFOS or non-EFOS. \par

%
\paragraph{ANN training} 
%
%

The training of the ANN is carried out from the following steps. All the RFCAs
previously identified as definitive EFOS are divided into two sets, one with
$2,981$ of the RFCAs, referred to as training set, and another one with $745$ EFOS,
the so-called test set. The same quantity of unknown RFCAs is added to the test set.
One million unknown RFCAs are added to the training set and then the $2,981$
definitive EFOS are added until we obtain the same quantity of unknown RFCAs,
ending up with a set of $2,000,000$ RFCAs. Thus, both sets will be made up of 50\%
of the data from definitive EFOS, corresponding to the randomly selected EFOS
income-type CFDI records, and 50\% of the unknown RFCAs randomly selected from
the total population. Data of the associated CFDI is obtained for each RFCA.
These are the data that are supplied to the ANN and from which the
ANN is trained by adjusting internal parameters. Following the training
process, the ANN is presented the set of test data that it has never seen
previously, to evaluate its performance. \par
\paragraph{Additional variables considered} 
In addition to incorporating the quantitative variables mentioned in
section~\ref{sec:data}, we attempted to incorporate categorical data
such as the type and situation of the contributor, the situation-description,
the status of the contributor, the start date of operations, the sector and
federal entity. We also considered incorporating data related to 
interactive networks (see section~\ref{sec:redes}), such as the degree of
output and input, betweenness, closeness, stress, radiality and page rank.
However, all the ANN trained with these variables performed equally or worse
than the ANN that only used CFDI data.

\subsubsection{Performance evaluation} 
\newcommand{\VP}{\textrm{TP}}
\newcommand{\TP}{\textrm{TP}}
\newcommand{\FP}{\textrm{FP}}
\newcommand{\FN}{\textrm{FN}}
\newcommand{\VN}{\textrm{TN}}
\newcommand{\TN}{\textrm{TN}}
We used the F1-score~\cite{Rijsbergen:1979:IR} as a measure to evaluate the competence of the
trained model. The F1-score is obtained by calculating the precision harmonic
mean and the recall. Precision is the proportion of relevant instances
correctly classified out of all the instances that the model believes are
relevant. If TP are the true positives and FP the false positives, precision
would be given by $\VP/(\VP+\FP)$ (see Table \ref{tab:MatrizConfusion}). Precision answers the question:
{\it How many of the selected RFCAs are actually EFOS?} Recall is the proportion of
the incorrectly classified relevant instances out of all the actually
relevant instances, $\VP/(\VP+\FN)$, where FN are the false negatives,  that
answers the question of all RFCAs that are actually EFOS: {\it How many were
correctly classified?} The harmonic mean is defined as the value obtained when
the number of values in the dataset is divided by the sum of its reciprocals.
It is a type of mean generally used for numbers that represent a ratio or
proportion (like the precision and recall) as it equalizes the weight of each
datapoint.  An F1-score attains its best value at 1 (perfect precision and
recall) and the worst at 0. Table \ref{tab:MatrizConfusion} shows a way of separating the
classifications made by the neuronal network to allow their evaluation. \par

\begin{table} [th!]
    \centering
    \begin{tabular}{ cc|c|c }
            &   \multicolumn{3}{c}{{Predicted class }}\\
            &   & P                     & N \\
        \cline{2-4}
        \multirow{2}{*}{{Real class}}
            & P & True positives (TP)   & False negatives (FN)\\
        \cline{2-4}
            & N & False positives (FP)  & True negatives (TN) \\
        \cline{2-4}
    \end{tabular}
    \caption{\label{tab:MatrizConfusion} 
Confusion matrix for binary classification. The true positives (TP) are the
examples that the model correctly classified as EFOS. The false negatives (FN)
are the examples that the model classified as non-EFOS, but that are actually
EFOS. The true negatives (TN) are examples that the model classified as
non-EFOS and have not been previously classified as EFOS. The false positives
(FP) are examples that the model classified as EFOS, but that were not
previously detected as such
%
%
}
\end{table} 


For example, if we take 500 definitive EFOS and 500 unknown ones, and we feed
them to our network, we find that $\TP = 448$, $\FN = 52$, $\TN = 416$ and  $\FP = 84$.
Therefore the precision was $0.845$, the recall was $0.896$, and we obtained an
F1-score of $0.87$. If we make the calculation with 1000 alleged EFOS, we
obtain $TP = 881$, $FN = 119$ ($TN = FP = 0$ by definition), so the precision is
1, while the recall is $0.881$. We thus obtain an F1-score of 0.94.  \par

The RFCAs in the set of ``alleged'' show the same behavior that the model
indentified by training with the set of ``definitive'', and it ends up
identifying 88\% as EFOS. \par

\begin{figure}[th!] 
    \centering
\includegraphics[width=\textwidth]{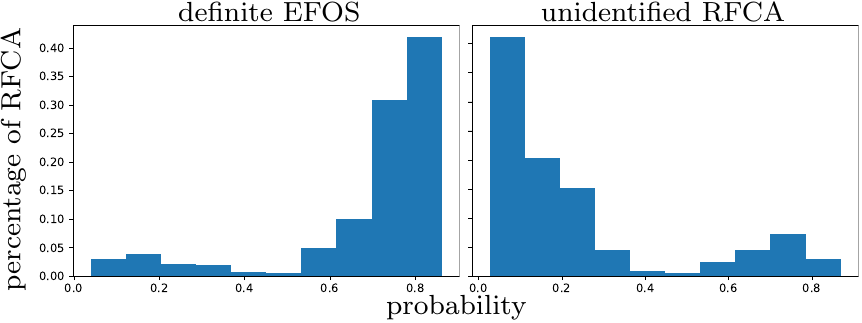}
     \caption{
Histograms of the probabilities assigned by the neural network to different
sets of RFCAs. (left) We observe that the network correctly
assigns most of them a high probability of being EFOS. (right) 
We observe a bimodal distribution in which there is a considerable
percentage of RFCAs that are assigned a high probability of being EFOS.
}
     \label{fig:hists_prob}
\end{figure}

In Figure \ref{fig:hists_prob} we observe that the model is sure of its
decision most of the time (i.e. ends with a very high or very low probability, with
a bimodal distribution). Additionally, in the probability distribution of the
non-identified RFCAs, there is a percentage that the model is classifying with
high probability (i.e., the model is sure that they are EFOS) but has not
previously classified them as EFOS.  \par

One of the greatest challenges in neural networks is to interpret what the
network is learning from the data. It is not only important to develop a solid
solution with great predictive power; it is also of interest to understand the
functioning of the developed model, i.e., which variables are the most
relevant, the presence of correlations, possible causal relationships, etc.
To deepen our understanding of the results, we applied two techniques to determine
the more relevant variables that we detail below.    

The first technique is based on hypothetical analysis or simulation, and is
used to measure the relative influence of the input variables on the model
results. In particular, to measure the importance of the variables, we took a
sample of our data $X$ and calculated the model predictions once $Y$ is trained.
Then, for each variable $x_i$ we cause a perturbation of this variable (and only
this variable) via a normal random distribution centered on 0 with a scale 20\%
of the variable mean, and calculate the prediction $Y_i$. We  measure the
effect exerted by this perturbation by calculating the difference of the
quadratic root mean between the original output $Y$ and the perturbed $Y_i$. A
larger mean square root difference means that the variable is ``more
important''. Table
\ref{tab:pert_var} (left) shows the five variables of greatest importance
for the neuronal network.

The second technique consists of the analysis of principal components, a
statistical technique to convert high-dimensional to low dimensional data by
selecting the most important characteristics that capture most of the
information about the dataset. Characteristics are selected as a function of
the variation they cause on the output. The characteristic that causes the
highest variance is the first component, the characteristic responsible for the
second highest variance is considered the second principal component, and so
on. It is important to mention that the principal components are not correlated
with each other. The importance of each variable is reflected in the magnitude
of the corresponding values in vectors that characterize a linear
transformation (greater magnitude indicates greater importance).
Table~\ref{tab:pert_var} (right)
lists the five variables that best characterize the dataset based on the first
principal component, which contributes 99\% of the variance. The variable
magnitudes are normalized so that the sum of squares is equal to 1.

\begin{table}[th!]
    \centering
    \begin{tabular}{  c | c }
         Variable & Perturbation effect\\
        \hline
         Sub-active& 0.2099 \\
         Active total& 0.1813 \\
         Total amount after active & 0.1419 \\
         IVA  after active & 0.1083 \\
         Cancelled amount & 0.0748 \\
        \hline
    \end{tabular}
\quad
    \begin{tabular}{ c | c }
         Variable & Magnitude \\
        \hline
         Active total & 0.74925125 \\
         Sub-active & 0.64598303 \\
         Total amount after active & 0.10326791 \\
         IVA  after active & 0.1032678 \\
         Cancelled amount & 0.0000125 \\
        \hline
    \end{tabular}
    \caption{\label{tab:pert_var}
(left) Effect of the perturbation on the probability assigned by the neural
network; (right) importance of the variables based on the absolute value of the
magnitude of the first principal component used to characterize the dataset.
}
\end{table} 

\subsubsection{Model results } 

The ANN efficiently classifies the identified EFOS that it has been presented
with, and using the trained model, we classify as
``suspicious'' the unknown RFCAs to which the ANN
assigns a greater probability of behaving similarly to published EFOS.  The ANN classified $149,921$ unknown RFCAs, corresponding to
1.98\% of the total, as suspicious, using the threshold probability ($> 0.8$).


\subsection{Random Forest} \label{subsec:randomForest}

As a second method of classification, we use the automatic technique named
{\it Random Forest} (RF). Techniques of automatic classification, including RF,
detect groups of elements with similar statistical patterns in an available
dataset, 
and from the knowledge acquired, make decisions about the membership
of new elements in these groups. In our case, we consider the characteristics
of EFOS published by SAT and we compare them to unknown RFCAs.
 
A RF is constructed by randomly combining different {\it decision trees} in
order to obtain results robust to noise sources inherent to the algorithm. A
decision tree is a mathematical algorithm made up of a set of questions ordered
and connected to each other through their responses (i.e., the formulation of a
question depends on the answer to a preceding question). These questions
involve the variables or characteristics of the data utilized. In constructing
a decision tree each node represents one of the questions and each fork depends
on its answer. Thus, in finishing the construction of a decision tree 
we can
follow a path determined by questions and answers, finally answering the main question:
how likely is this RFCA to be an EFOS?

In statistical models like RF it is necessary to maintain a balance between
measures such as the {\it variance} (variability in the prediction of models for
different elements) and {\it bias} (the difference between actual and predicted
value). To achieve this balance, an effective technique is the combination of
various models, e.g., the combination of decision trees to form a RF. In this
way, each decision tree issues a classification (i.e., a probability of
suspecting it is an EFOS associated to a RFCA) and the final result of the RF
is the most probable classification among all the trees constructed. One of the
tasks to resolve in constructing a RF is to find the optimum number of decision
trees used to determine the combination that generates the final solution. 

In our case, the RF technique is considered adequate since it offers the
following advantages: 
\begin{itemize}
\item Data preparation is minimal. It is only necessary to rely on a dataset
where each element to classify, in this case each RFCA, is unique and has a
fixed number of characteristics associated with each one of the classes
involved, in this case definitive or unknown.
\item It can handle a large number of variables without discriminating any one of them.
\item It has been demonstrated that it is one of the methods with highest precision among the classification algorithms~\cite{breiman2001random}.
\item It performs well with large-volume databases (which applies to the
present case study) The result of the RF is a number between 0 and 1 for each
evaluated RFCA, which will be interpreted as the probability that each unknown
RFCA is a potential EFOS. 
\end{itemize}
\subsubsection{Data preparation} 

For the implementation of the RF algorithm, the information by issuer is
initially grouped, since this analysis is focused on classification of issuing
RFCAs. As a result, a unique record for each issuing RFCA is obtained for each
of the 48 months considered.

Subsequently, by means of a technique called
{\it undersampling}~\cite{liu2008exploratory} a
balanced sample of unknown RFCA and definitive EFOS is generated. This
technique seeks to determine the optimal number of RFCAs that will allow
obtaining a balanced sample of the data (i.e., has the same quantity of
unknowns and definitives) as well as a representative one (i.e., that will
capture the characteristics of the whole population with the number of RFCAs
selected. This process yields a sample with 1561 definitive EFOS and 1561
unknown RFCAs. The sample obtained so far is the baseline dataset used for
implementation of the RF algorithm.

As part of the data preparation phase, two
independent treatments are applied to the previously generated sample:
\begin{enumerate}
\item The data were analyzed to determine what type of data transformation is
viable for each of the sample variables. The family of {\it box cox} transformations
was used to improve the normality of the data and equalize the variance in
order to improve the algorithm’s performance~\cite{osborne2010improving}.
\item
Principal components analysis (PCA) was used. This consists in reduction of the
dimensionality of the dataset by unifying existing variables to create new
ones. This method is recommended to improve the performance of the algorithms
in question~\cite{wold1987principal}.
\end{enumerate}

\subsubsection{Model construction}  

Using the RF algorithm three models were constructed that correspond to the
following scenarios and use the sample generated in the previous section:
\begin{itemize}
\item First scenario: implementation of the RF algorithm without
transformation 
\item Second scenario: implementation of the RF algorithm using the data sample
to which PCA was applied.
\item Third scenario: implementation of the RF algorithm using the data sample
to which the {\it box cox} transformation 
was applied.
\end{itemize}
For each of the above scenarios, training of the RF algorithm aims to find the
optimum number of constituent decision trees. This is achieved by performing
iterations of the algorithm, modifying the number of trees used, and
determining when the error generated stabilizes at a minimum value. It was
concluded that the optimal number of decision tress was 100.

\subsubsection{Performance evaluation} 

The following measures were used to evaluate the above scenarios:
\begin{itemize}
\item
Receiver operating characteristic (ROC) curve: is a performance measure with
values between 0 and 1; the higher the value, the better the performance is
considered. An ROC curve is constructed using the information from two
characteristics: the sensitivity (possibility of appropriately classifying
a positive individual, in this case a definitive EFOS) and the specificity
(possibility of appropriately classifying a negative individual, in this case
an unknown RFCA that is not actually a definitive RFCA)~\cite{martinez2003curva}.
\item 
Error: is a penalty measure. The closer it is to 0, the better it is
considered. The error quantifies the part of the model that is making a mistake
in classifying the RFCAs, and in the case of  %
a RF is obtained through a
combination of the error generated by each one of the individual trees, as well
the correlation between them~\cite{breiman2001random}.
\end{itemize}

 \begin{table} [th!]
    \centering
    \begin{tabular}{ l|c|c }
        {Escenario} & {ROC} & {Error} \\
        \hline
         Random forest & 0.912 & 0.164 \\
         Random forest plus principal components& 0.886 & 0.161 \\
         Random forest plus variable transformation & 0.893 & 0.157 \\
        \hline
    \end{tabular}
    \caption{\label{tab:machine_learning}
Comparison of performance measures for the different ways in with the input
data were transformed.} 
    	\label{tab:BAmodelos}
\end{table} 
As shown in Table~\ref{tab:BAmodelos}, even though there is an
improvement in the performance for the first scenario, error reduction is
favored; therefore the model selected was the one that included the Box Cox
transformation of data. This was the model used in the following steps.

\begin{table} [th!]
    \centering
    \begin{tabular}{ c|c|c|c|c }
\multicolumn{1}{c}{}& \multicolumn{4}{c}{Years with activity} \\\hline
Years classified as  EFOS & 1 year & 2 years & 3 years & 4 years \\
        \hline
0 & 17\% (133)& 5\%  (56) & 3\%  (11)&  6\% (8)  \\
1 & 83\% (631)& 13\% (143)& 6\%  (24)&  4\% (6)  \\
2 &           & 82\% (893)& 17\% (71)& 11\% (16) \\
3 &           &           & 74\%(307)& 26\% (37) \\
4 &           &           &          & 53\% (77) \\
        \hline
    \end{tabular}
    \caption{\label{tab:efos_anual} 
We study the performance of the RF algorithm over the years. We
considered the definitive EFOS, separated by the number of years with activity
(columns). In the different rows, we consider the number of years in which the
algorithm classifies RFCAs as EFOS; thus, a definitive EFOS should be detected
by the algorithm in at least one of the years of activity. For example, RF
erroneously classified 3\% out of the total definitive EFOS with activity
reported over 3 years, which corresponds to 11 definitive EFOS.
}
\end{table} 
An additional validation was conducted considering the selected model,
which consisted of classifying the definitive EFOS using the model (which we
know {\it a priori} should have have a high probability), and observing the outcome.
A cut-off point of 0.8 was established, i.e., if the risk index is greater or
equal than 0.8, the classified RFCA is considered an EFOS,
otherwise it is not.
Additionally, the years of activity of each definitive EFOS were considered for
the final diagnosis, e.g., if it was active for two years, the two
qualifications are considered, and so on. Results shown in Table \ref{tab:efos_anual} were
obtained by developing the above, where it can be observed that close to 92\%
of the definitive EFOS are being correctly classified by the algorithm, and the
error is only 8\%.

 \begin{table} [th!]
    \centering
    \begin{tabular}{ l|c|c }
        {Classification} & {Frequency} & {Percentage} \\
        \hline
         EFOS & 1,908  & 79\% \\
         No EFOS & 505 & 21\% \\
        \hline
    \end{tabular}
    \caption{\label{tab:calificacion}
Classifying the different types of contributors.
}
\end{table} 

Combining the above results, those RFCAs that in all years of activity
were detected by the model were classified as possible EFOS, and as non-EFOS if
not.  Table \ref{tab:calificacion} shows that out of all definitive EFOS, only
505 were classified as non-EFOS, which means that they are the only ones about
which the algorithm is completely wrong. This behavior is considered normal due
to the possibility that the EFOS may have engaged in illegal activities only in
some years.

\subsubsection{Results} 
Using the model developed and validated in the previous sections (third
scenario), we take four groups of unknown RFCAs (one per study year) and
determine the risk index. Note that if the RFCA has been active for more than
one year, it will have a different index each year.

Based on the previous results the following groups were defined for all unknown
RFCAs:
\begin{itemize}
\item
Suspicious: all those unknown RFCAs that in each year of activity have a risk
index greater or equal than $0.8$.
\item
Not suspicious: all unknown RFCAs that in at least one of the years of activity
have a risk index $< 0.8$. 
\end{itemize}
Based on these definitions the algorithm classified 7,438,448 RFCAs (98.3\%) as
not suspicious and 128,227 RFCAs (1.7\%) as suspicious of being EFOS.


\subsection{Complex network approach}\label{sec:redes} 

In this section we describe the way in which we define interaction networks between EFOS and
unclassified RFCA based on the emission and reception of CFDIs. We also describe the analysis we
perform on the topology of the network and the roles EFOS and the rest of the RFCA play in them.
This analysis allowed us to build the metrics used to define suspect evaders in the interaction
networks.

\subsubsection{Interaction network definition} 

\begin{figure} [th!] 
    \centering
    \includegraphics[width=0.5\linewidth]{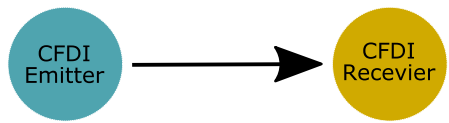}
    \caption{
        Directed links in the network correspond to a CFDI emitted between RFCA. This digital
        receipts (CFDI) can represent incomes and outcomes.
    }
    \label{fig:def_enlace}
\end{figure}

The taxpayers activity records allows us to define interaction networks between them in which nodes
correspond to taxpayers (identified by their RFCA) and are classified in one of three categories:
definitive EFOS (those evaders already identified by the authorities), alleged EFOS (suspect evaders
identified by the authorities), and unclassified RFCA. Links in the network correspond to directed
transactions between taxpayers, which as CFDI themselves, can represent either income or outcome
transactions (see Fig.~\ref{fig:def_enlace}).

The topology of these networks represent the relationships between groups of taxpayers, which we
assume reflect some of the association patterns and mechanisms EFOS and other RFCA have used
for their practices and which we use to identify suspect evaders.

With the available data we construct yearly and monthly interaction networks. On one hand, the year
timescale allow us to identify a set of RFCA with whom EFOS interact more regularly. On the other
hand, we have identified on the month timescale, that the amounts associated with transactions
between EFOS occur more frequently inside an interval we have termed \textit{EFOS activity regime},
which correspond to higher amounts than those observed in transactions between unclassified RFCA. We
build the interactions networks between RFCA, only taking into account the transactions with amounts
inside this interval and characterize their topology and structure.
\subsubsection{Yearly interaction networks}\label{ssect:redes_anuales} 

We first consider the interaction network built only from the income CFDI emissions and receptions
of the nodes associated to EFOS with at least 10 transactions in a year. This restriction selects
the nodes that interact more frequently during a year (at least once a month which), according to
the homophily principle in social networks~\cite{aiello2012friendship, mcpherson2001birds,
currarini2016simple,asikainen2020cumulative}, correspond to nodes which are more similar between them.

\begin{figure} [th!] 
    {\includegraphics[width=0.48\linewidth]{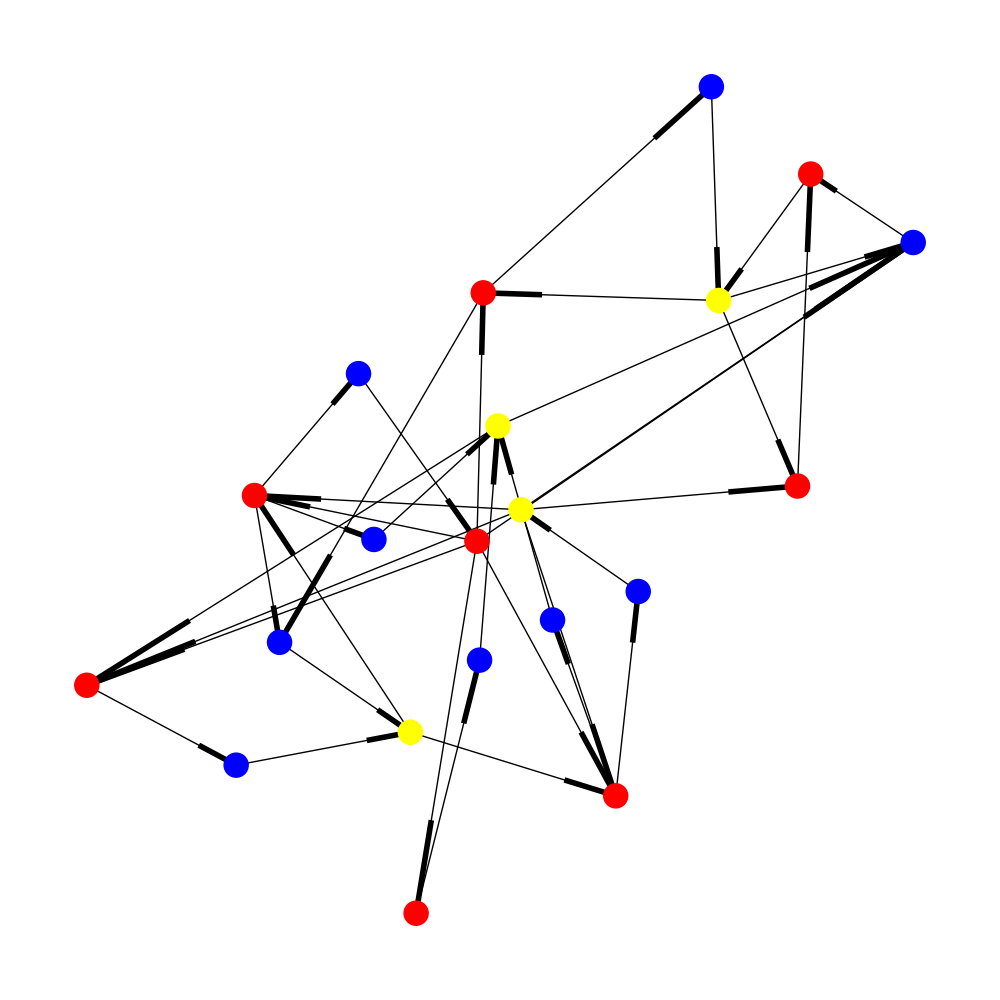}}
    {\includegraphics[width=0.48\linewidth]{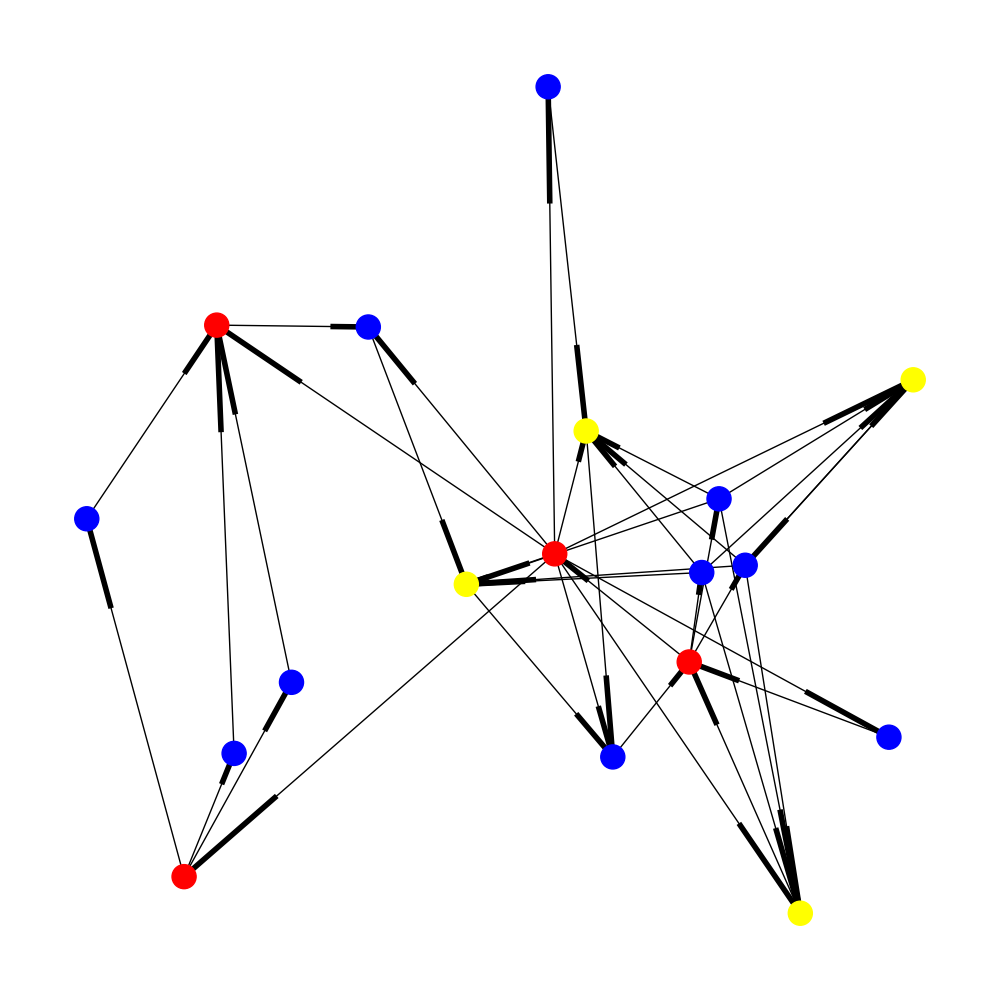}}

     \caption{ Examples of the strongly connected components observed in interaction networks at
     year timescale, the left panel corresponds to 2015 and the right panel to 2016. Red nodes
 correspond to confirmed EFOS, yellow to suspect EFOS and blue nodes to unclassified RFCA.  }
 \label{fig:ejemplos_coms} \end{figure} 

We identify the strongly connected components (SCC) in these networks with organized sets of
taxpayers which can be related to anomalous financial activity. We show the largest SCC observed in
2015 and 2016 in Fig.~\ref{fig:ejemplos_coms}. Remembering that links in the network correspond to
transactions, the presence of these structures imply a circular flow of goods and services, which
being related to nodes associated to EFOS, is possible that these sets of nodes carry out tax
evasion practices such as the exchange of receipts of simulated transactions, which suggests that
the unclassified RFCA in the SCC might be suspect of carrying out the same practices.

\subsubsection{Monthly interaction networks}\label{ssec:nivel_oper} 

In this section we consider the interaction networks between taxpayers at the month timescale.
Unlike the yearly interaction networks where we only took into account the emissions and receptions
of the nodes associated with EFOS, in this case we consider the transactions between all three kinds
of nodes (EFOS, alleged EFOS and unclassified RFCA). Nonetheless, as the whole set of transactions
is huge, we need to define criteria to filter transactions to reduce the network into a manageable
size.

\begin{figure}[th!] 
    \centering
    \includegraphics[width=0.495\linewidth]{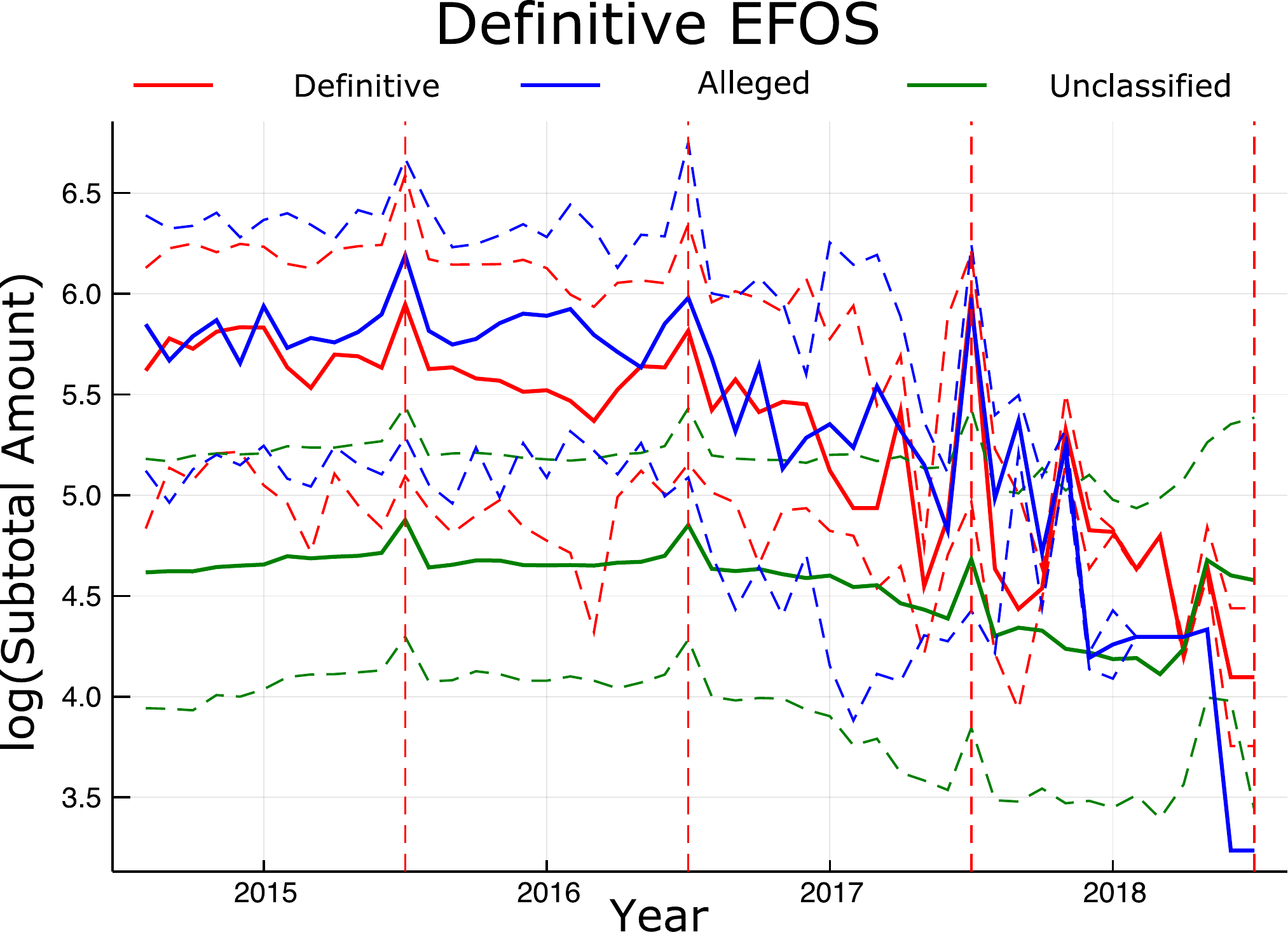}
    \includegraphics[width=0.495\linewidth]{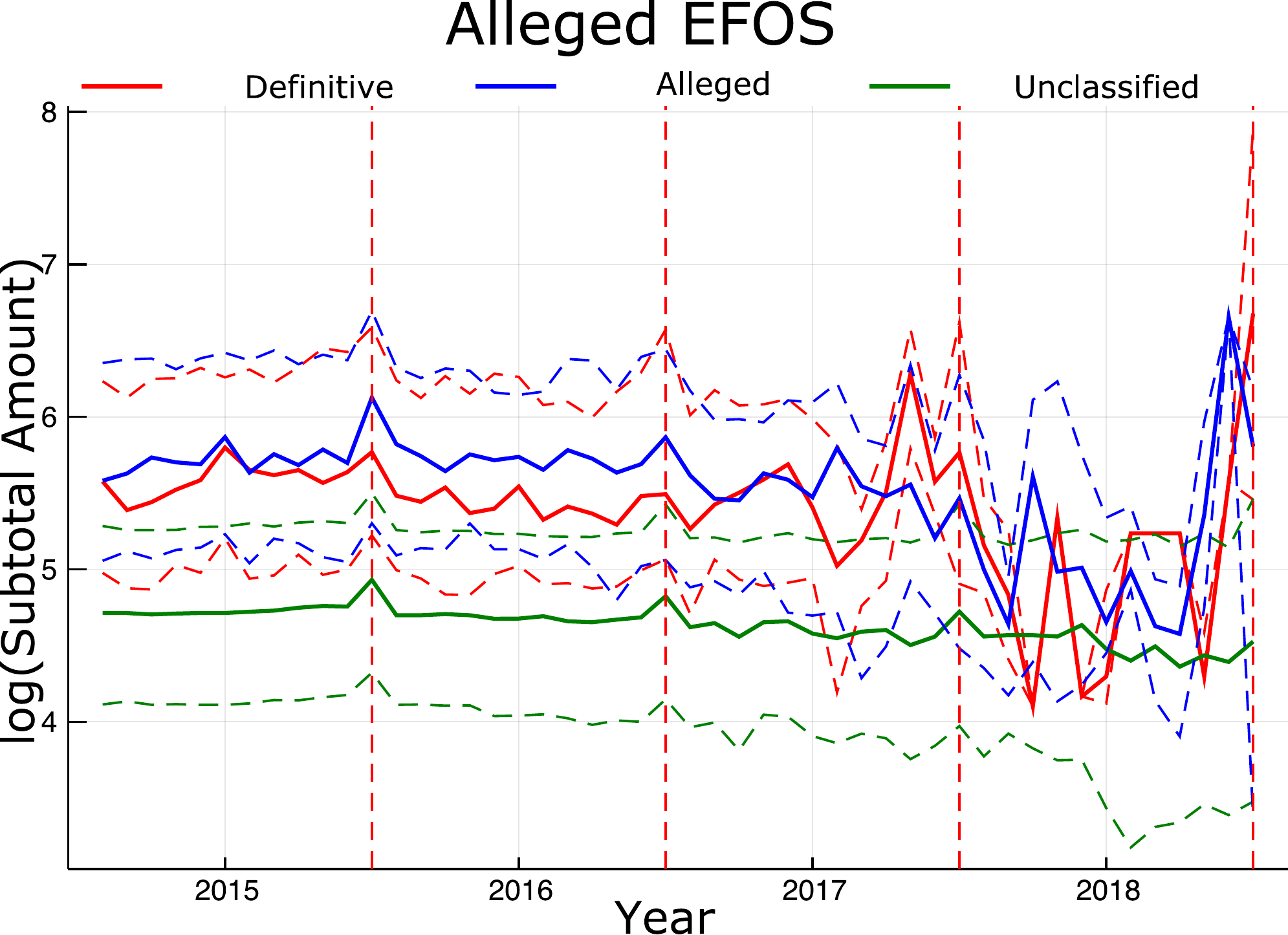}
    \caption{
        Time behaviour of the logarithm of the subtotal of the transactions associated to
        emissions from EFOS (left panel) and alleged EFOS (right panel), the vertical dashed lines
        indicate the month of December of each year. Solid lines show the median and the dashed
        lines the interquartile range of the distribution. It can be seen that the transactions
        between EFOS, either definitive or alleged, correspond to amounts around tens of thousands
        and millions of pesos. We define this range of amounts as the EFOS activity regime, which we
        use to filter the links in the interaction networks.
    }\label{fig:emision_dif}
\end{figure} 

To this end, we obtain the distribution of the subtotal amounts before taxes of the transactions
emitted by EFOS (for both definitive and suspect) to the remaining types of nodes. As can be seen in
Fig.~\ref{fig:emision_dif}, the median of the distribution changes over time, showing an increase
towards the end of the year. It is to be noted that the amounts of the transactions between EFOS are
higher than those of the transactions between EFOS and unclassified RFCA, which suggests that EFOS
make selective emissions whether the receiver of the transaction is an EFOS or arbitrary RFCA. We
define the EFOS activity regime, as the amounts interval defined by the interquartile ranges of the
amounts distribution obtained from the EFOS transactions. We only take into account links in the
monthly interaction networks whose amount lies inside the EFOS activity regime. Once we have
selected the relevant links in the monthly interaction networks between taxpayers, we obtain the
largest SCC of the network, which e.g. for 2015 consists of 635,588 nodes.

We define the reach of the \(i\)-th node in the network at distance \(d\), \(r_i(d)\), as the
fraction of nodes in the SCC up to a distance \(d\) from the node, i.e.
\begin{subequations}
    \begin{equation}
        \omega_i = \{ j \;|\; d_{ij} \leq d\},
    \end{equation}
    \begin{equation}
        r_i(d) = \frac{|\omega_i|}{N},
    \end{equation}
\end{subequations}
where, \(d_{ij}\) is the length of the shortest path length between nodes \(i\) and \(j\),
\(|\omega_i|\) is the number of elements in \(\omega_i\), and \(N\) is the number of nodes in the
SCC.

If we now consider a set of nodes \(\Omega\), we can calculate the mean reach of the set of nodes at
distance \(d\), \(\langle R(d) \rangle_{\Omega}\), by
\begin{equation}
    \langle R(d) \rangle_{\Omega} = \frac{1}{|\Omega|} \sum_{i \in \Omega} r_i(d),
\end{equation}
where \(\Omega\) represents either the set of nodes that correspond to EFOS or unclassified RFCA.

\begin{figure} [th!] 
    \centering
    \includegraphics[width=0.49\linewidth]{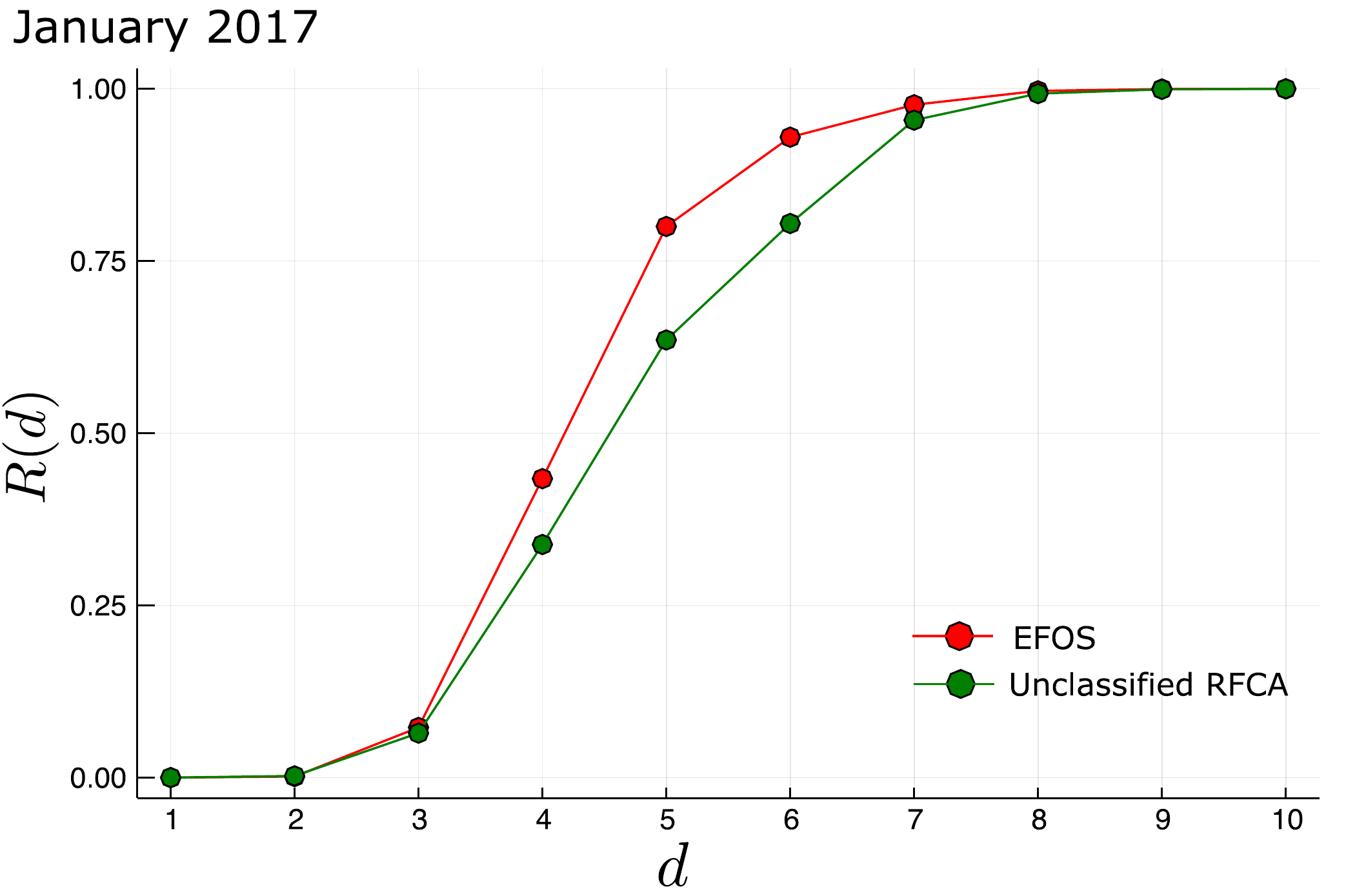}
    \includegraphics[width=0.49\linewidth]{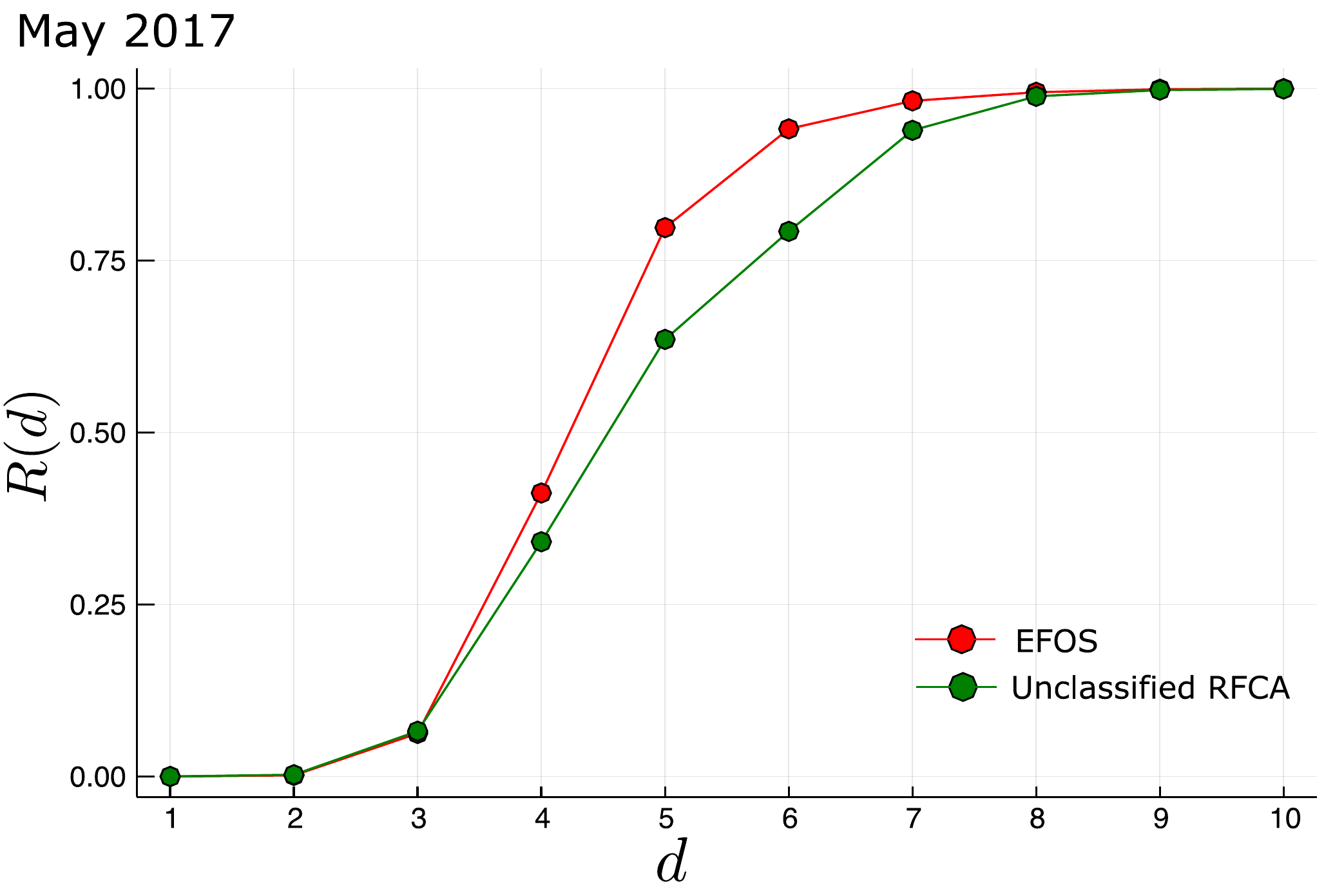}
    \caption{
        Mean reach, \(R(d)\), as a function of distance \(d\), for nodes that correspond to EFOS and
        to unclassified RFCA. \(R(d)\) is higher for EFOS than for unclassified RFCA, which suggests
        that EFOS are more efficient to distribute their transactions in the network. Data
        correspond to January 2017 (left panel) and May 2017 (right panel).
    }
    \label{fig:reach_month}
\end{figure} 

\begin{figure} [th!]
    \centering
    \includegraphics[width=0.49\linewidth]{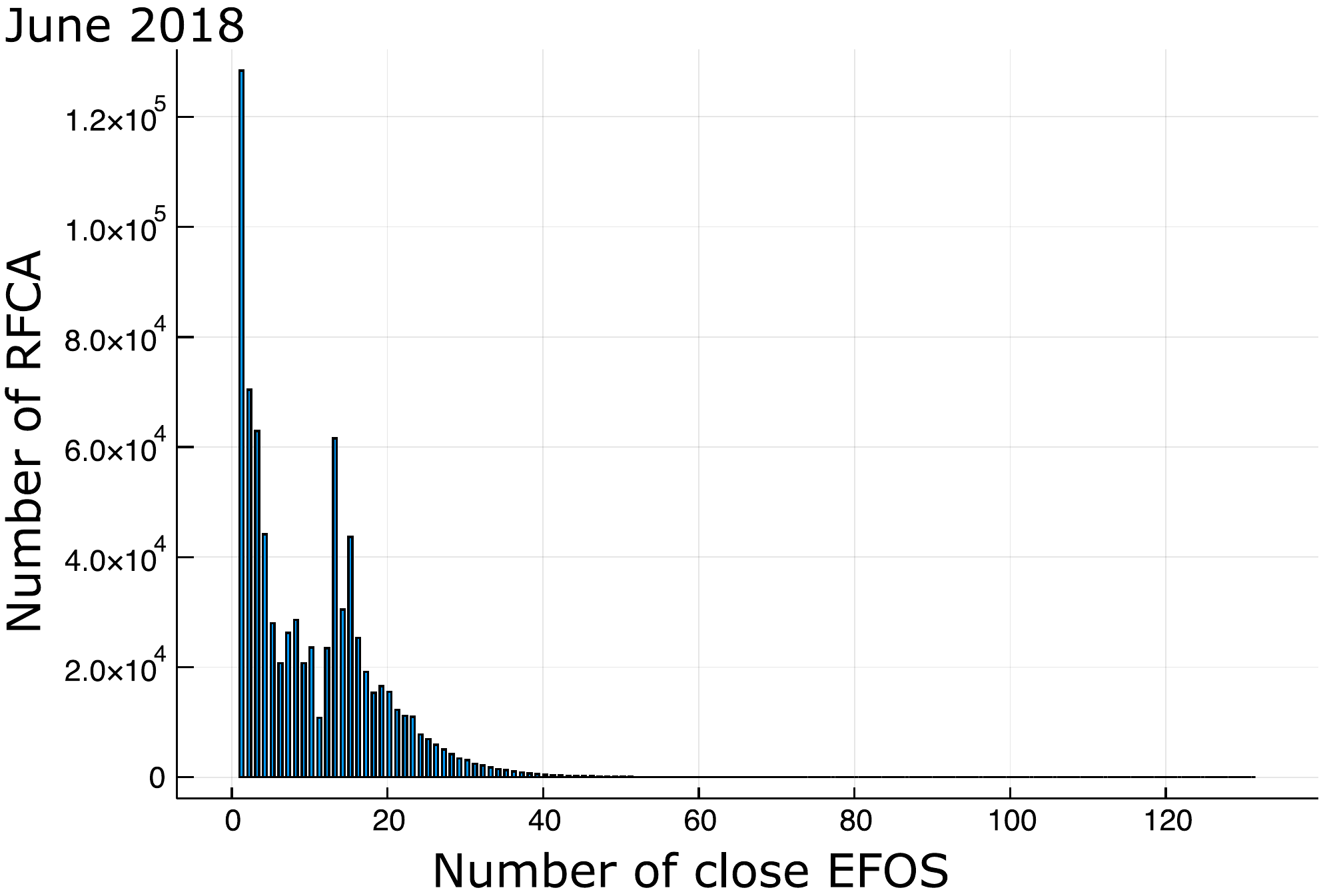}
    \includegraphics[width=0.49\linewidth]{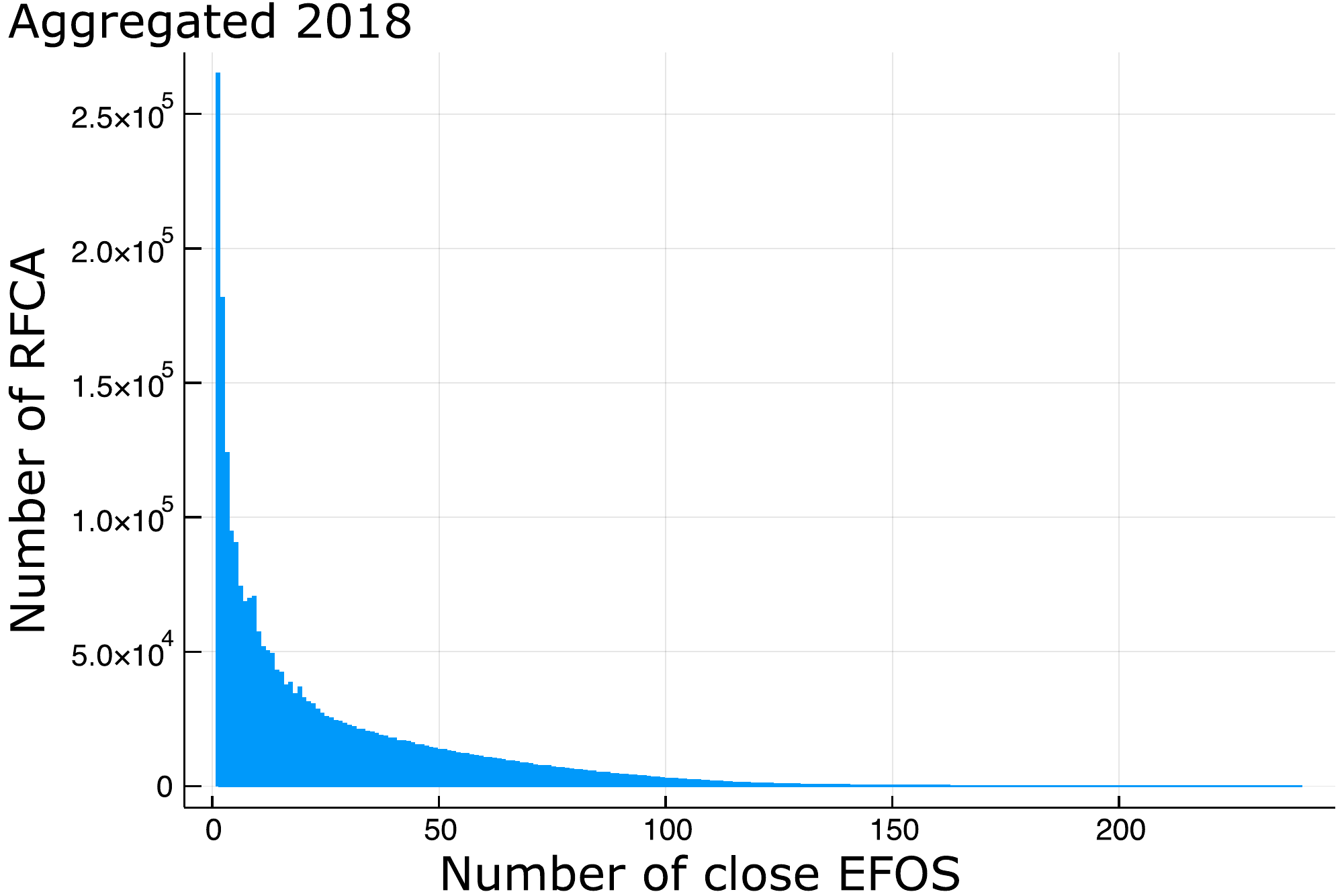}
    \caption{
        Nearest EFOS to unclassified RFCA distribution for the monthly interaction networks (left
        panel) and one year's aggregate (rigt panel). The number of close EFOS is a collusion
        indicator of unclassified RFCA in the EFOS operation networks. As can be seen in both
        panels, there are unclassified RFCA close to a large number of EFOS in both timescales.
    }\label{fig:cercania}
\end{figure}

The topology of the network is such that, as can be seen in Fig.~\ref{fig:reach_month}, the reach of
EFOS is higher for distances \(3 < d < 7\) than for the rest of the nodes. This suggests that EFOS
are more efficient to distribute their transactions in the network, which can be related to
mechanisms aimed to limit the traceability of their transactions. Following the reach behavior, we
define the set of nearest neighbors of a node as the set of nodes at \(d_{ij} < 3\). We plot the
distribution of close EFOS, for both a month and one year's aggregates. From unclassified RFCA in
Fig.~\ref{fig:cercania}, we see there are cases in which unclassified RFCA are close to
more than 100 EFOS in one month.


We use the number of close EFOS to unclassified RFCA in the interaction network as an indicator of
their collusion in the EFOS operation networks, so that we can assume that an arbitrary RFCA close
to a large number of EFOS, is more susceptible to take part in the same corrupt practices as EFOS,
when compared to RFCA which are not as close to EFOS in the network.

We define for each one of the identified suspect RFCA, the \textit{EFOS proximity index},
\(\sigma_i(y)\), for the \(i\)-th node in the network for the year \(y\), as que quotient of the
total number of EFOS at \(d_{ij} < 3\) during the year, and the number of months these EFOS were close
to the RFCA, i.e.
\begin{equation}
    \sigma_i(y) = \frac{\mathrm{Close\;EFOS\;in\;} y}{ \mathrm{Number\;of\;months\;they\;were\;close} }.
\end{equation}
Note that the denominator can be less than 12, as there can be months in which the RFCA
wasn't close to any of the EFOS in the network. As the number of active EFOS in the network changes
over time, we normalize the EFOS proximity index, which we express by \(\hat{\sigma}_i\), with
respect to the maximum observed during the year, i.e.
\begin{equation}
    \hat{\sigma}_i(y) = \frac{\sigma_i(y)}{\max(\sigma_i(y))},
\end{equation}
where \(\hat{\sigma}_i(y) \in [0,1]\), this allows us to define a threshold to be used in all
periods, \(\theta_\sigma(y)\), which through the condition \(\hat{\sigma}_i(y) \geq \theta_\sigma(y)
\) with \(\theta_\sigma(y) \approx 1\), allows us to select the more colluded suspect RFCAs for each
year.


\begin{figure}[th!] 
    \centering

    \subfloat[\label{fig:indice_y_15}]{\includegraphics[width=0.49\linewidth]{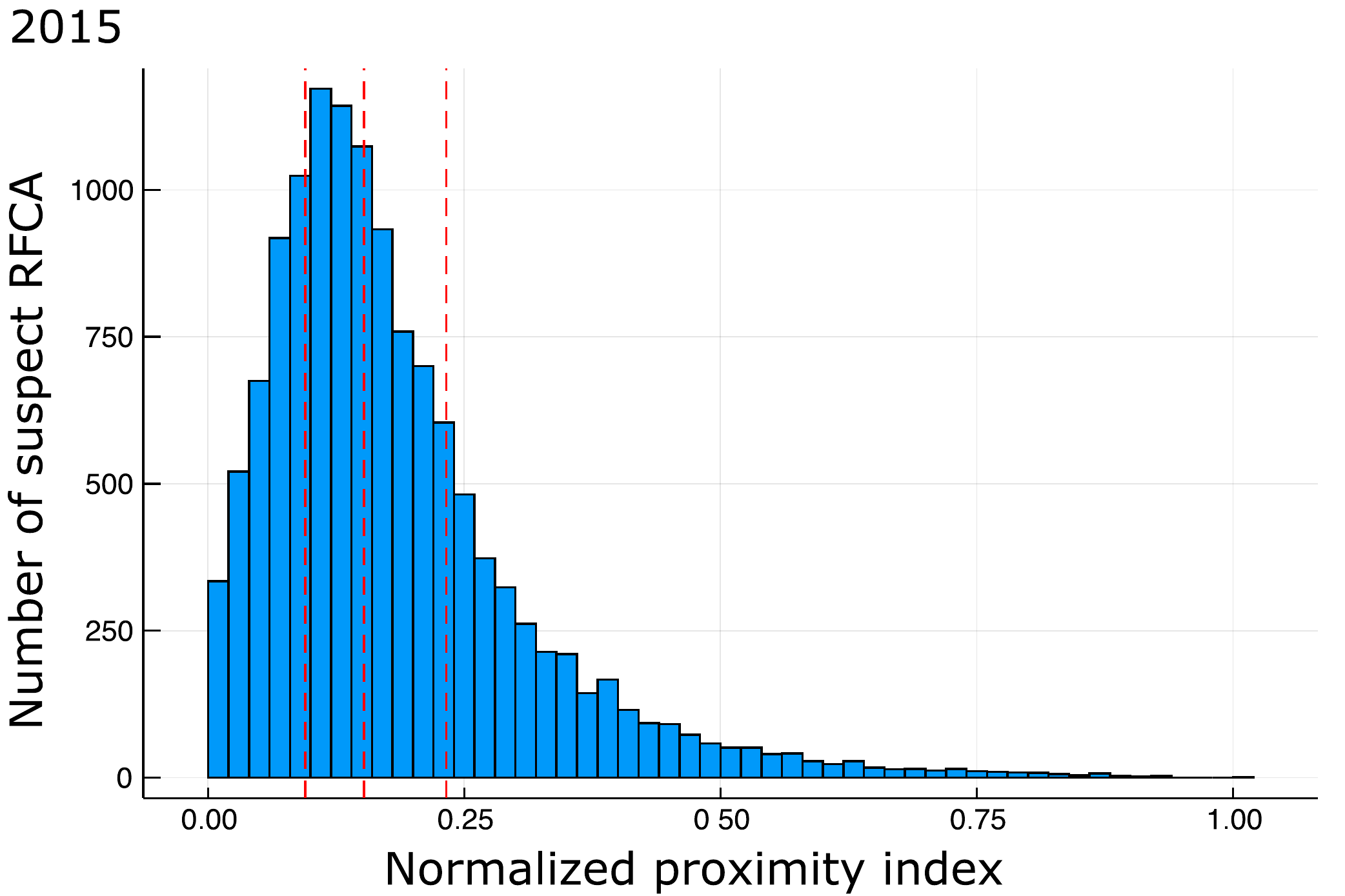}}
    \subfloat[\label{fig:indice_y_16}]{\includegraphics[width=0.49\linewidth]{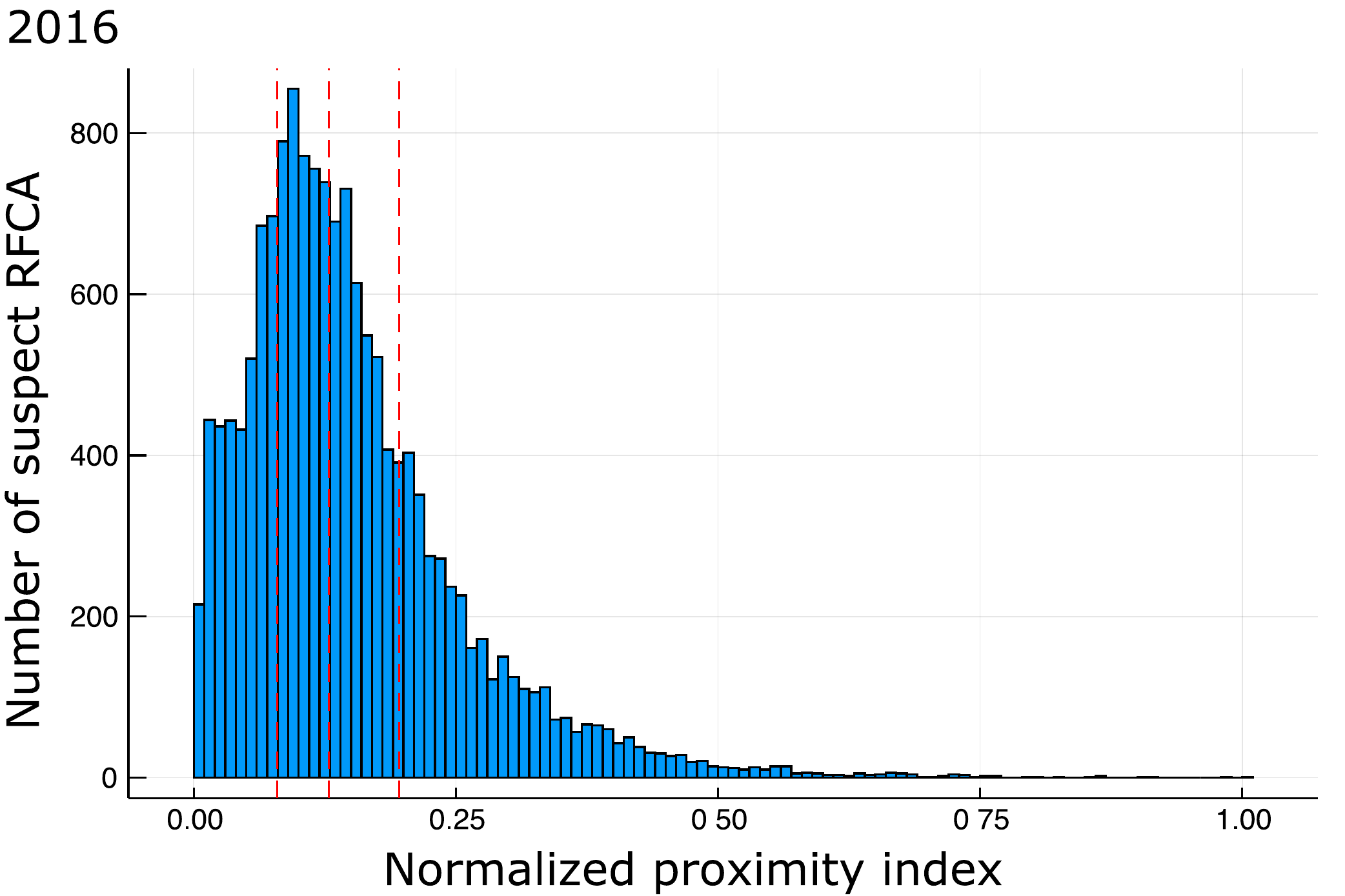}}

    \caption{
        Normalized proximity index to EFOS distribution for suspect RFCA identified by the ANN and
        Random Forest models.  This index is used as an extra validation criteria of the suspect
        list obtained from the Machine Learning models using characteristics observed in the
        interaction networks. We show the results for (a) 2015 and (b) 2016. Dashed lines represent
        the 25, 50 and 75\% quartiles of the proximity index distribution.
    }\label{fig:cercania_sosps}
\end{figure} 
 

The description we've made of both yearly and monthly interaction networks between taxpayers,
allowed us to identify features of the EFOS organization mechanisms and the local structure of the
network around them, such as: their organization in small operation networks related to closed flows
of CFDIs between them, and selective CFDI emissions between EFOS of much larger amounts than the
transactions they make with unclassified RFCAs, which suggests that EFOS operate inside an activity
regime defined by the amounts of their transactions (see Fig.~\ref{fig:emision_dif}). We have been
able as well to quantify, by means of the reach of nodes in the network and the proximity index, the
collusion of unclassified RFCAs inside the EFOS operation networks. These results suggest that
complex networks analysis provides useful techniques with ample potential to describe and
characterize corruption networks and operational mechanisms, which are yet to be fully explored.


\subsection{Merging the classification methods} 
\label{sec:integracion}

We obtain a list of suspect RFCAs from each Machine Learning classification method (ANN and Random
Forest). It is to be noted that the training set for both of these methods, consisted of examples
from the evaders already identified by the authorities, therefore, there is an inherent and implicit
bias in our methods towards the mechanisms and assumptions used by the authorities to identify
evaders. This bias in unavoidable due to the data we had available and it is important to take into
account further methods to make a wider characterization of these and other evasion mechanisms.

We report the number of suspect RFCAs identified by each method in Table~\ref{tab:sosps_metodos}. The
list obtained from the ANN corresponds to those RFCAs with an probability of belonging to the same
class as EFOS \(> 0.8\). We use the same threshold for the probability assigned by the Random Forest
model. The intersection of both suspect lists is contains 43,650 RFCA, which we consider to have a
higher probability of being suspect evaders, as they were identified independently by two different
methods.

\begin{table}[th!] 
    \centering
    \begin{tabular}{ c|c|c }
        Method & Classified as suspect & Unclassified \\
        \hline
        Artificial Neural Network & 149,921 & 7,416,754 \\
        Random Forest             & 128,227 & 7,438,448  \\
        \hline
    \end{tabular}

    \caption{
        Number of suspect RFCA identified by each classification method.
    }\label{tab:sosps_metodos}
\end{table} 

\subsubsection{Suspect RFCA comparison with EFOS} 

\begin{figure}[th!] 
    \centering

    \subfloat[\label{fig:box_mt_act}]{\includegraphics[width=0.45\linewidth]{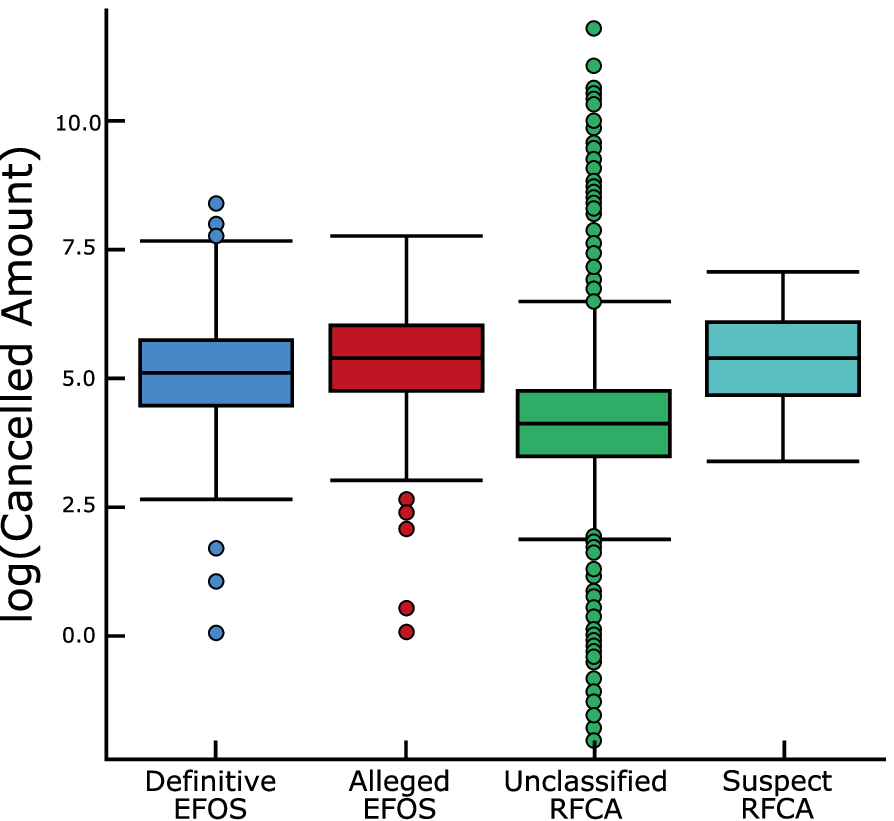}}
    \hspace{1cm}
    \subfloat[\label{fig:box_sub_act}]{\includegraphics[width=0.45\linewidth]{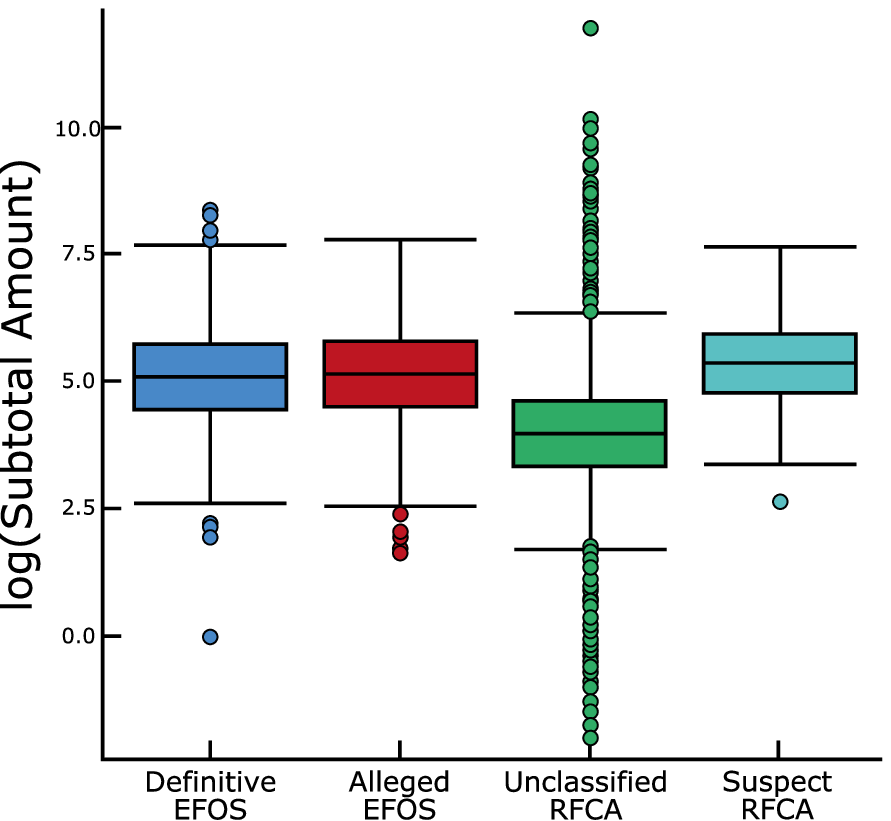}}

    \caption{
        Boxplots for the distributions of the (a) logarithm of the total cancelled amount and (b)
        the active subtotal amount of the CFDI emitted by EFOS, suspect and unclassified RFCA.  It
        can be seen that the distributions for EFOS and suspect RFCA are more similar between each
        other than when being compared with the distributions of unclassified RFCA.
    }\label{fig:boxPlots}
\end{figure}

\begin{figure}[th!] 
    \centering

    \subfloat[\label{fig:no_quant_mt_can}]{\includegraphics[width=0.45\linewidth]{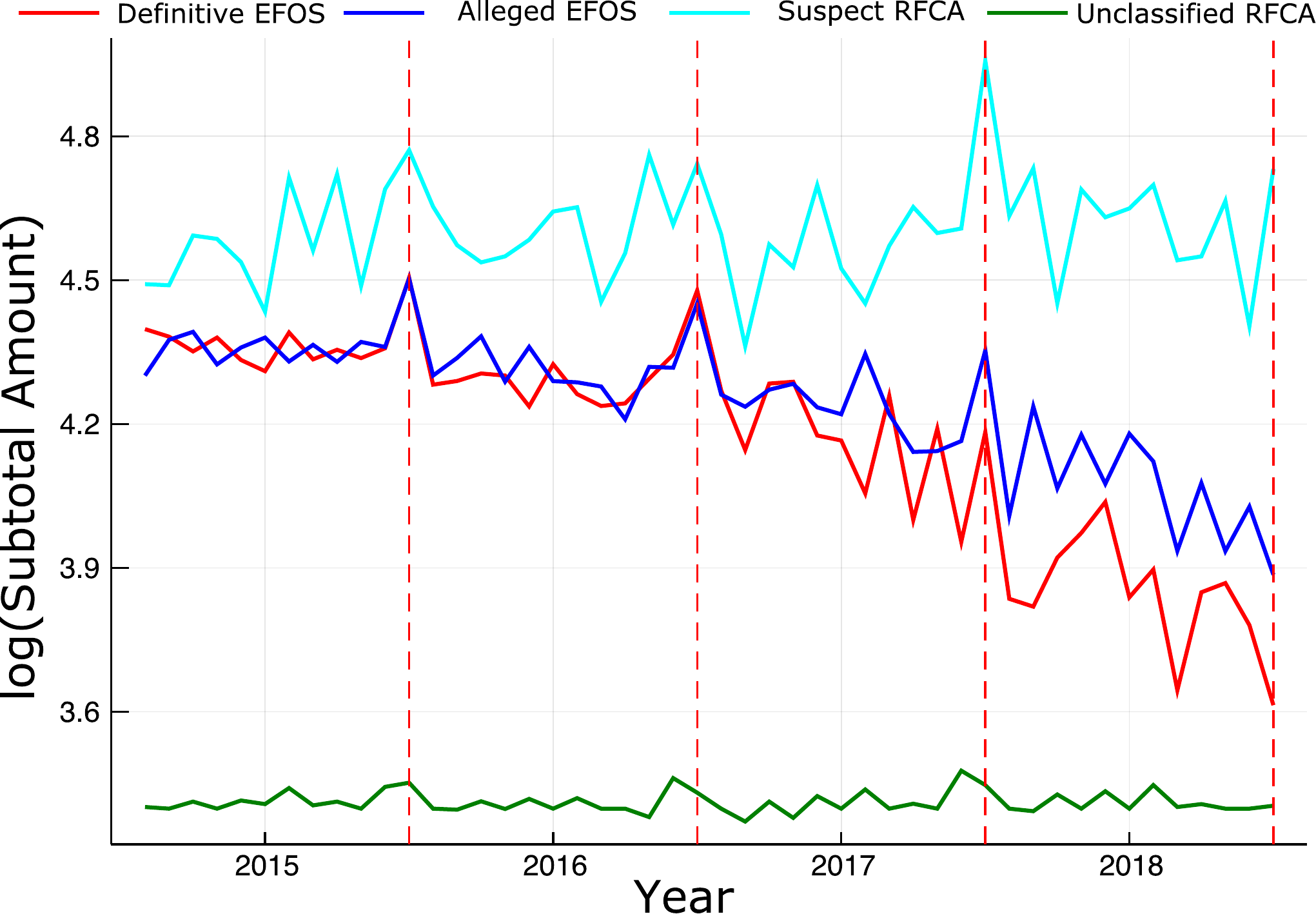}}
    \hspace{0.04\linewidth}
    \subfloat[\label{fig:no_quant_sub_act}]{\includegraphics[width=0.45\linewidth]{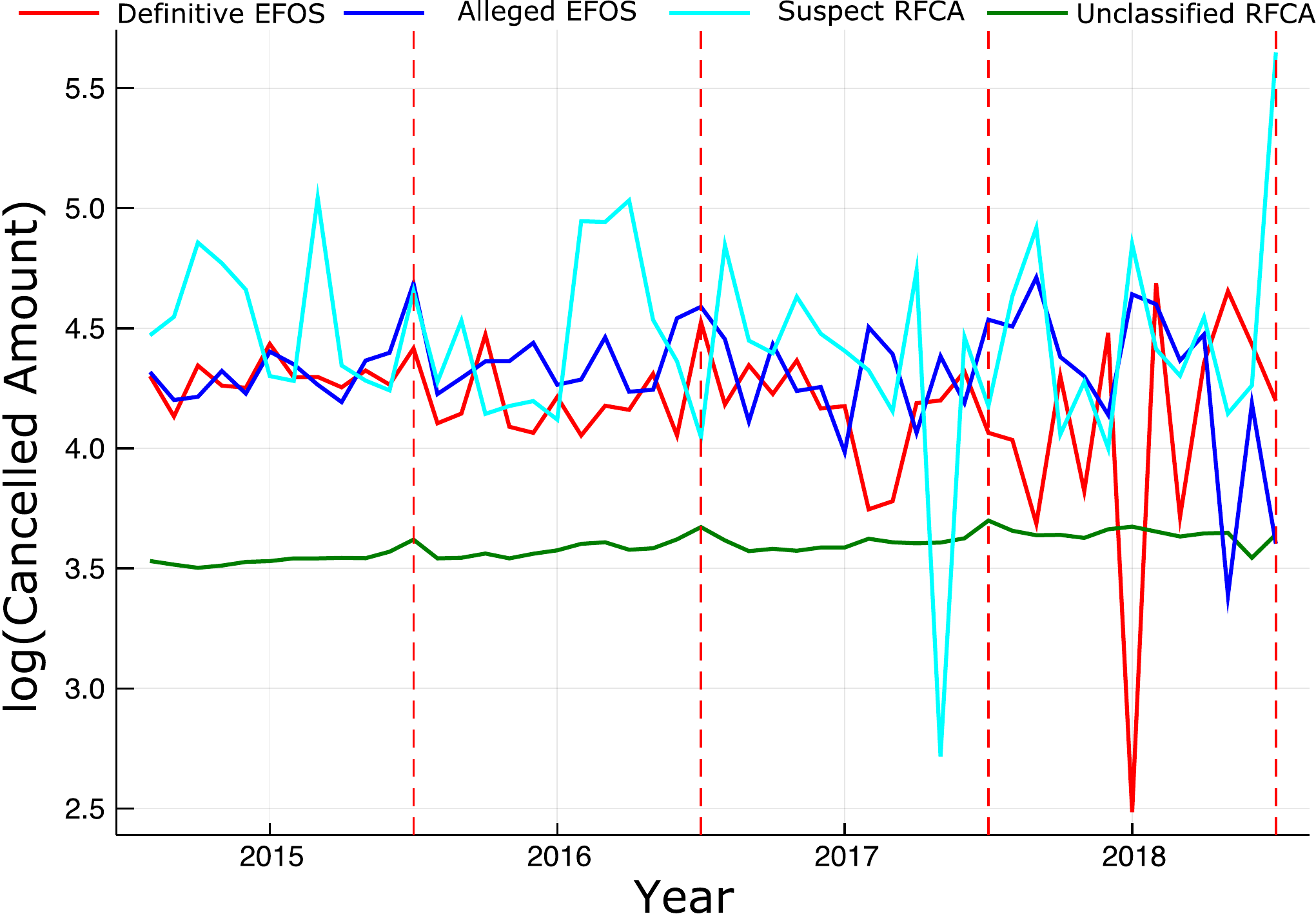}}

    \caption{
        Temporal behavior of the (a) Active subtotal amount and (b) Cancelled total amount for each
        one of the groups of RFCAs: definitive EFOS (blue), alleged EFOS (red), unclassified RFCAs
        (green) and suspect RFCAs (cyan). The vertical dashed lines correspond to December for each
        of the years studied. It can be seen that the EFOS and suspect RFCAs behavior differentiates
        from unclassified RFCAs.
    }\label{fig:sops_trend}
\end{figure}

To compare the features of the suspect RFCAs identified by our classification methods with those of
EFOS, we compute the distributions of the active and canceled amounts associated to their CFDI
emissions. Active amounts correspond to the emitted and processed invoices, and canceled amounts
correspond to those invoices that were canceled and not processed. The cancellation of CFDIs can be
done freely and independently by taxpayers without an explicit authorization. As can be seen in
Fig.~\ref{fig:boxPlots}, the distributions of these two variables are very similar for EFOS and the
suspect RFCAs identified by our methods, while for unclassified RFCA, the distributions of these
variables are different. Furthermore, in Fig.~\ref{fig:sops_trend}, we show the time behavior of the
active and canceled amounts of the CFDIs emitted by the different RFCA groups (definitive EFOS,
alleged EFOS and unclassified RFCAs). It can be seen that the difference between suspect and
unclassified RFCAs is consistent for the whole analysis period, which sets these two groups (EFOS
and suspect RFCA) apart from unclassified RFCA.


\subsubsection{Number of close EFOS to suspect RFCA} 
    
\begin{figure}[th!] 
    \centering

    \subfloat[\label{fig:sosps_num_efos_y_15}]{\includegraphics[width=0.49\linewidth]{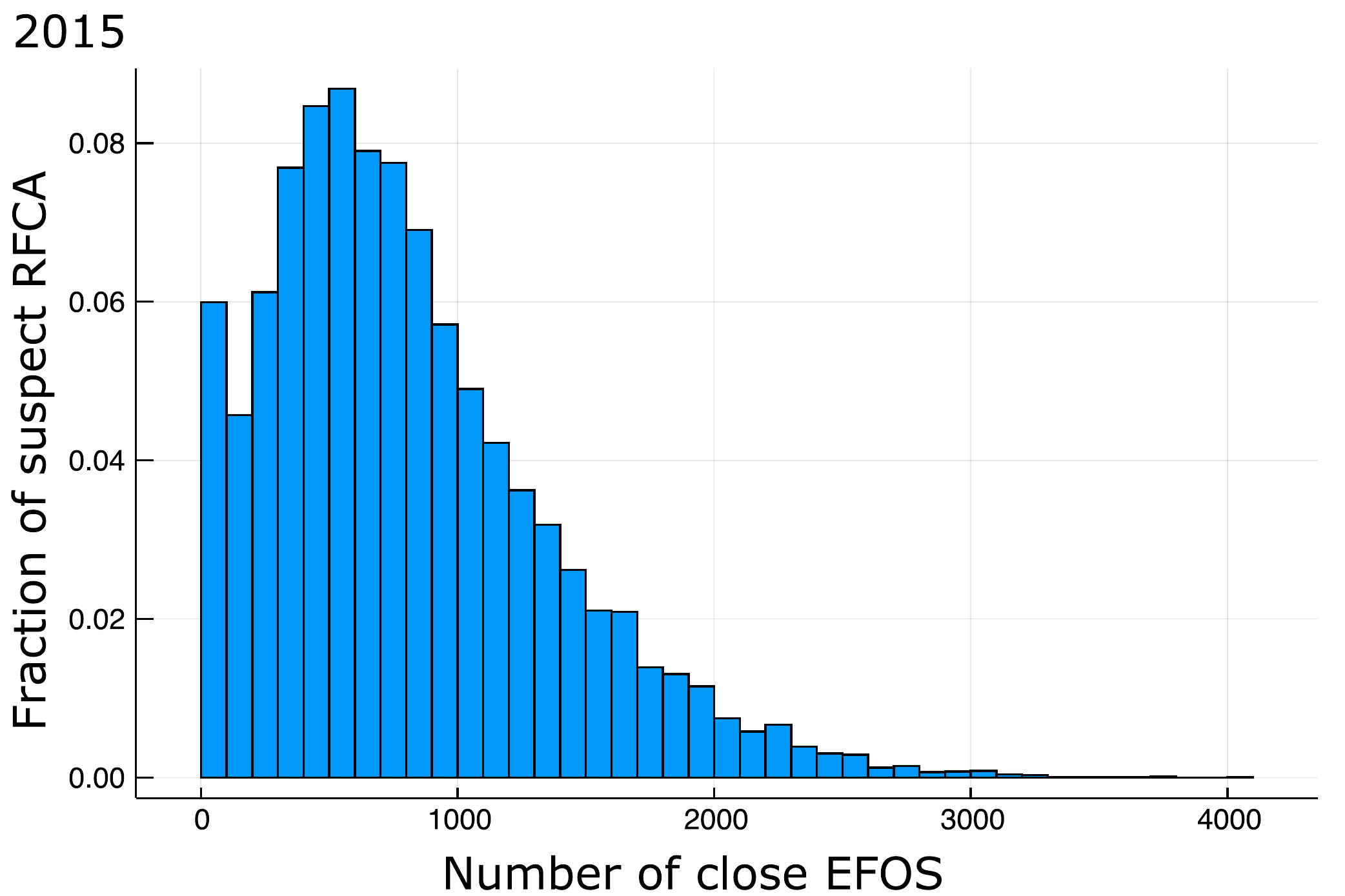}}
    \subfloat[\label{fig:sosps_num_efos_y_16}]{\includegraphics[width=0.49\linewidth]{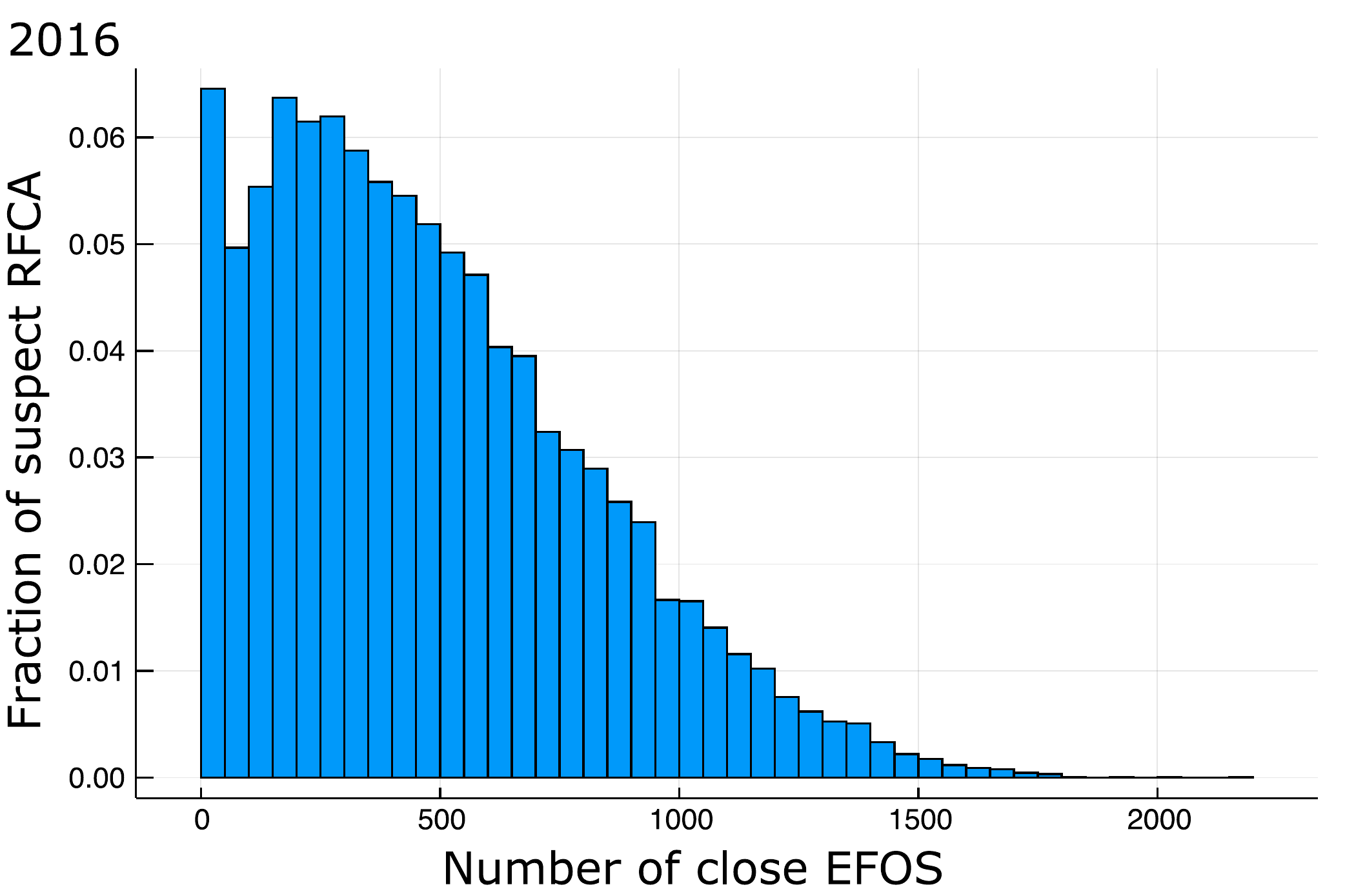}}

    \caption{
        Total of close EFOS the suspect RFCAs identified by our classification methods for (a) 2015
        and (b) 2016. It can be seen that a suspect RFCA are close to a large number
        of EFOS, which indicates that these RFCAs are immersed into the EFOS operation networks.
    }\label{fig:efos_cercanos_sosps}
\end{figure}

As it was discussed in Section~\ref{sec:redes}, the EFOS reach in the interaction networks allows us
to identify the number of close EFOS to unclassified RFCA in the network (where close means at a
distance \(d \leq 3\)) and use it to select those RFCAs more immersed in the EFOS operation
networks.  Now, if we look at the suspect RFCAs identified by the classification methods, and
compute the number of close EFOS in a whole year, we observe that they are close to a large number
of EFOS (see Fig.~\ref{fig:efos_cercanos_sosps}), which suggests that the suspect RFCAs in the
intersection of the lists obtained from both classification method are colluded with the EFOS
identified by the authorities and gives us confidence about our methods. It is to be noted that the
closeness to EFOS wasn't part of the set of features used by the classification methods (ANN and
Random Forest) to identify suspect RFCAs, as they were based only on CFDI data.


As we show in Table~\ref{tab:frac_pers}, 81.52\% of the suspect RFCA are legal persons, which
suggests that most part of the emitted CFDI linked with allegedly simulated operations is made
between companies or businesses. This can be due to the fact that, in this case the legal
responsibility of possible illicit operations, falls to a legal entity and not a natural person. In
table~\ref{tab:frac_pers}, we also show that 91.15\% of the suspect RFCA were active at the moment
of the elaboration of the study, and less than 1\% of them had a cancelled or suspended status,
which shows that most of the suspect RFCA are economically active, and thus susceptible of being
investigated.


\begin{table} [th!]
    \centering
    \begin{tabular}{ c|c }
            Taxpayer type & Percentage of suspect RFCAs \\
        \hline
            Legal        & 81.52\% \\
            Natural      & 10.22\% \\
            Without info & 8.3\%   \\
        \hline
    \end{tabular}
    \hspace{0.25cm}
    \begin{tabular}{ c|c }
        Status & Percentage of suspect RFCAs \\
        \hline
        Active       & 91.15\% \\
        Cancelled    & 0.13\%  \\
        Suspended    & 0.46\%  \\
        Without info & 8.3\%   \\
        \hline
    \end{tabular}

    \caption{
        Type of taxpayer and status of the suspect RFCA identified by the classification methods.
        Most of them were active legal taxpayers at the moment, which made them susceptible of being
        investigated by the authorities.
    }\label{tab:frac_pers}
\end{table}


Using the data contained in the CFDI, we define the potential tax collection associated to an
arbitrary RFCA, \(\mathrm{rec_{VAT}}\phi_i\), as the difference between the total nominal tax
obtained by the income CFDI emitted during the year, \(\mathrm{VAT_{Nom}}_i\), and the payed tax
reported in their tax statements, \(\mathrm{VAT_{Payed}}_i\), i.e.
\begin{equation}
    \mathrm{rec_{VAT}}\phi_i = \sum \mathrm{VAT_{Nom}}_i - \mathrm{VAT_{Payed}}_i,
\end{equation}
so that, we define the total nominal tax collection that correspond to the set of suspect RFCA,
\(\mathrm{REC_{IVA}}\phi \) as:
\begin{equation}
    \mathrm{REC_{VAT}}\phi = \sum_i \mathrm{rec_{VAT}}\phi_i.
    \label{eq:def_rec}
\end{equation}

We use the income CFDI emitted by the identified suspect RFCA between the years 2015 and 2018, and
their yearly tax statements, which include the total VAT payed by them. We believe that the
information of the emitted CFDI allows us to have a better description of the economical activity
and evasion mechanisms, because the income amounts declared in the tax statements is vulnerable of
manipulation, and may not reflect reality, mostly because we are dealing with taxpayers suspects of
simulating operations. A significative difference between income amounts from CFDI emissions and tax
statements could be a indicator of possible illicit practices.


\subsection{Yearly evasion estimates} 

As we discussed in section~\ref{sec:redes}, the number of close EFOS to a RFCA in the interaction
networks, is an indicator of their collusion level in the sub-networks associated with EFOS, so that
we can assume that a suspect RFCA close to a larger number of definitive and alleged EFOS, is much
more susceptible to conduct the same practices as EFOS. With this in mind, we calculate evasion
estimates for each year between 2015 and 2018, based on the suspect RFCA closest to EFOS in the
monthly interaction networks for each year, which correspond between 28\% and 38\% of all suspect
RFCA.


\begin{figure}[th!] 
    \centering
    \includegraphics[width=0.75\linewidth]{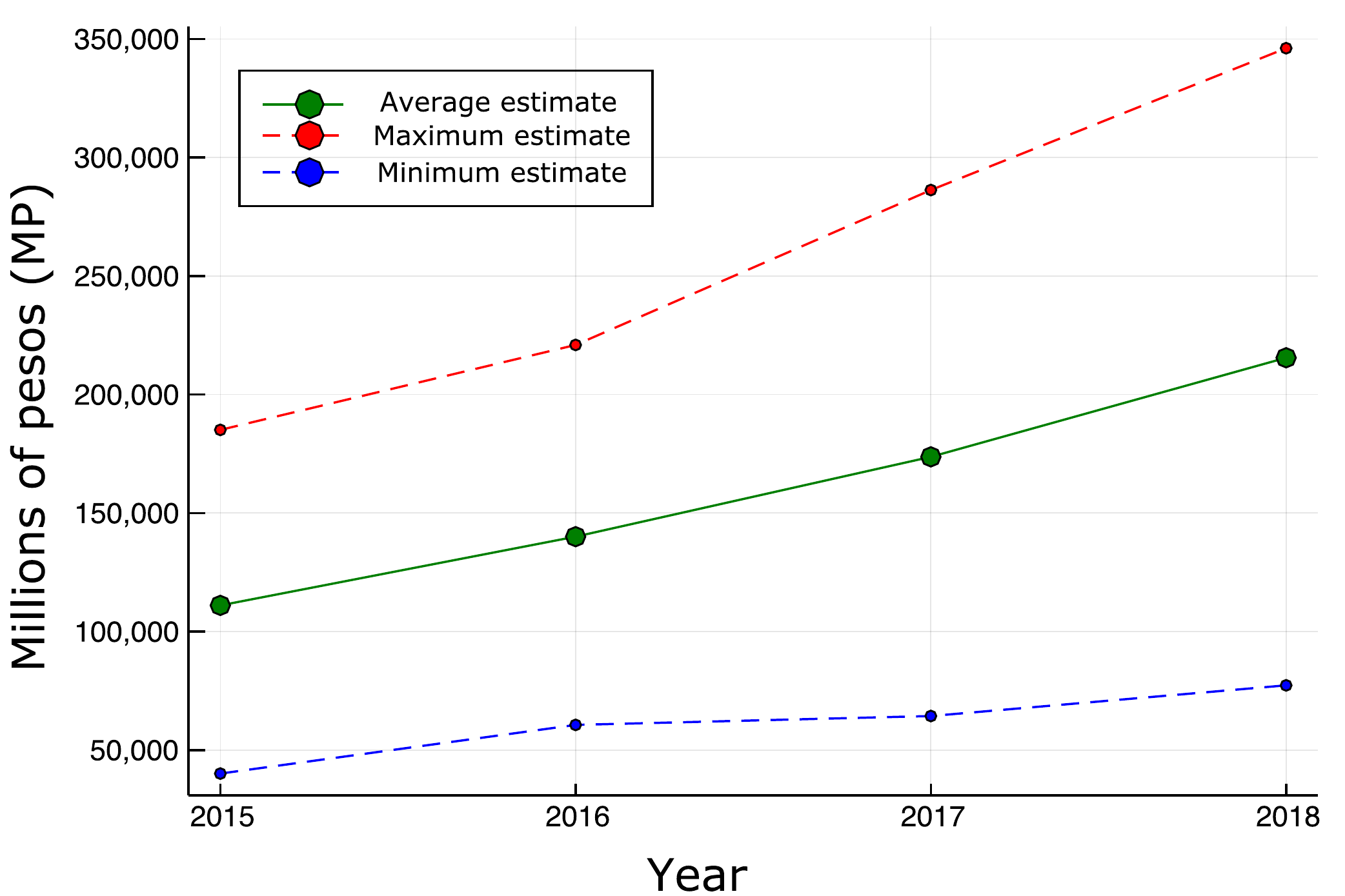}

    \begin{tabular}{ c | c | c | c  }
        \hline
        \hline\noalign{\smallskip}
        \multicolumn{4}{c}{VAT evation estimates (MP)} \\
        \hline\noalign{\smallskip}
        \hline
        Year & Minimum & Average & Maximum \\
        \hline
        2015 &  40,097.27 & 111,048.36 & 185,087.23 \\
        2016 &  60,626.86 & 140,041.13 & 220,922.03 \\
        2017 &  64,377.11 & 173,717.06 & 286,273.35 \\
        2018 &  77,318.59 & 215,518.71 & 346,106.32 \\
        \hline
        \hline\noalign{\smallskip}
        Year average & 60,604.96 & 135,081.31 & 259,597.23 \\
        \hline
        \hline
    \end{tabular}
    \hspace{1.5cm}
    \begin{tabular}{ c | c | c  }
        \hline
        \hline\noalign{\smallskip}
        \multicolumn{3}{c}{Unique suspect RFCA} \\
        \hline\noalign{\smallskip}
        \hline
        Year & Minimum estimate & Maximum estimate \\
        \hline
        2015 & 2,686 & 10,767 \\
        2016 & 3,132 & 12,510 \\
        2017 & 3,461 & 13,743 \\
        2018 & 3,541 & 14,080 \\
        \hline
        \hline\noalign{\smallskip}
        Total & 7,677 & 17,769 \\
        \hline
        \hline
    \end{tabular}

     \caption{
         Maximum and minimum VAT evasion estimates in millions of pesos (MP) associated to the
         emission of CFDI of potentially simulated operations from suspect RFCA in the period 2015
         -2018. We present as well the number of unique RFCA used for the calculations for each
         year. We estimate the number of unique suspect RFCA to be between 7,677 and 17,769.
     }\label{fig:montos_cotas}
\end{figure} 


The EFOS proximity index threshold, \(\theta_{\sigma}\), allows us to define minimum and maximum
values for VAT evasion estimates. The maximum estimated value corresponds to all identified suspect
RFCA, and minimum values to estimates based on suspect RFCA of the last quartile (top 25\%) of the
EFOS proximity index distribution, which correspond to 7,677 unique suspect RFCA with CFDI emissions
between 2015 and 2018, which corresponds to an average estimate of 60,604.96 millions of pesos per
year. The VAT evasion estimates in Fig.~\ref{fig:montos_cotas} should not be considered as final
values, as there might exist VAT evasion mechanisms different from the simulation of operations,
which we don't take into account in this work. Furthermore, it is important to mention that there is
an additional uncertainty because, it is not possible to determine precisely the fraction of
simulated and real transactions made by the suspect EFOS with the available information. Because of
this situation, we assume the 100\% of the emitted CFDI by suspect RFCA as simulated operations. A
follow up study focused in the traceability of the emitted CFDI could help determine the fraction
of simulated operations, which would allow us to make a more precise VAT evasion estimate.


\section{Discussion} \label{sec:discussion}

We have shown that it is possible to use tools from network science and machine learning to automatically identify patterns of tax evaders similar to those that humans already identified. This is promising, but should also be taken with caution. It is promising, because similar techniques could be applied in a broad variety of areas: money laundering, bribery practices, and other illegal activities. In principle, this would benefit society. However, one should also be aware of the limits of these methods. First, the patterns identified are based on those already known to humans. This means that different patterns of tax evasion will not be identified. Second, using these techniques to curb illegal activities would promote criminals to use alternative patterns that are not identified, so an arms race would ensue. The tools have a relevant potential, but by themselves, are not a final solution. Finally, there is always the probability of misidentifying honest citizens or companies for wrongdoers, simply because they have similar statistical patterns. Thus, these methods can be used to identify and prioritize \emph{potential} suspects from a huge pool, but the final decision has to be made by humans. 

In general, our work illustrates the potential of recent technology to solve different problems exploiting large data sets, computing power, and sophisticated statistical techniques. Still, the limits of this technology are yet to be defined properly, which has led to much hype and overconfidence. Also, ethical issues must be considered for the use of this technology. As more applications are developed, we can learn from them to better situate the benefits and risks of using network science, machine learning, and related techniques to address social problems.
\begin{acknowledgement}
This manuscript describes research associated with a project advising the Tax
Administration Service (SAT) of the Mexican federal government. The official
report of the project (in Spanish) is available in the SAT website at
\url{http://omawww.sat.gob.mx}. We thank Juan Pablo de Bottom, Alejandra
Cañizares Tello, Leonardo Ignacio Arroyo Trejo, and Aline Jacobo Serrano at
SAT, as well as Alejandro Frank Hoeflich, José Luis Mateos Trigos, Juan Claudio
Toledo Roy, Ollin Langle, Juan Antonio López Rivera, Eric Solís Montufar,
Octavio Zapata Fonseca, Romel Calero, José Luis Gordillo, and Ana Camila Baltar
Rodríguez at UNAM. G.I. acknowledges partial support from the Air Force Office
of Scientific Research under award number FA8655-20-1-7020, and by the EU H2020
ICT48 project Humane AI Net under contract 952026. C.P. and C.G. acknowledge
support by projects CONACyT 285754 and UNAM-PAPIIT IG100518, IG101421,
IN107919, and IV100120.

\end{acknowledgement}

%
%
\bibliographystyle{unsrt}
\bibliography{references,references_cp}
%

\end{document}